\documentclass[conference]{IEEEtran}
\usepackage{mathtools}
\usepackage{amsmath} 
\usepackage{amssymb}
\usepackage{amsfonts}
\usepackage{verbatim} 
\usepackage{bm,color,soul}
\usepackage{algorithm,algorithmic}
\usepackage{cite}
\usepackage{caption}
\usepackage{subfigure}
\usepackage{stfloats} 
\usepackage{graphicx} 
\usepackage{subeqnarray}
\usepackage{cases}
\usepackage{makecell} 
\usepackage{enumerate}
\usepackage{array}
\usepackage{url}
\usepackage{balance}

\newtheorem{prop}{Proposition}
\newtheorem{lem}{Lemma}

\newcommand{\rnum}[1]{\uppercase\expandafter{\romannumeral #1\relax}}
\ifCLASSINFOpdf
\else
\fi
\makeatletter
\renewcommand{\maketag@@@}[1]{\hbox{\m@th\normalsize\normalfont#1}}
\makeatother

\IEEEoverridecommandlockouts

\begin{document}
	%
	\title{Joint Transmit and Reflective Beamforming for Multi-Active-IRS-Assisted Cooperative Sensing \vspace{-3mm} }

	\author{\IEEEauthorblockN{Yuan Fang\IEEEauthorrefmark{1}, Xianghao Yu\IEEEauthorrefmark{1}, and Jie Xu\IEEEauthorrefmark{2}\\ \IEEEauthorblockA{\IEEEauthorrefmark{1}Department of Electrical Engineering, City University of Hong Kong \\
				 \IEEEauthorrefmark{2}SSE and FNii, The Chinese University of Hong Kong (Shenzhen)\\
				  Email: yuanfang@cityu.edu.hk, alex.yu@cityu.edu.hk, xujie@cuhk.edu.cn}\vspace{-0.5cm}	
		}
	}
	\maketitle
	\begin{abstract}
		This paper studies multi-active intelligent-reflecting-surface (IRS) cooperative sensing, in which multiple active IRSs are deployed in a distributed manner to help the base station (BS) provide multi-view sensing. We focus on the scenario where the sensing target is located in the non-line-of-sight (NLoS) area of the BS. Based on the received echo signal, the BS aims to estimate the target's direction-of-arrival (DoA) with respect to each IRS. In addition, we leverage active IRSs to overcome the severe path loss induced by  multi-hop reflections. Under this setup, we minimize the maximum Cram\'{e}r-Rao bound (CRB) among all IRSs by jointly optimizing the transmit beamforming at the BS and the reflective beamforming at the multiple IRSs, subject to the constraints on the maximum transmit power at the BS, as well as the maximum transmit power and the maximum power amplification gain at individual IRSs. To tackle the resulting highly non-convex max-CRB minimization problem, we propose an efficient algorithm based on alternating optimization, successive convex approximation, and semi-definite relaxation, to obtain a high-quality solution. Finally, numerical results are provided to verify the effectiveness of our proposed design and the benefits of active IRS-assisted sensing compared to the counterpart with passive IRSs.
	\end{abstract}
	

	%
	\IEEEpeerreviewmaketitle
	
	\vspace{-0.3cm}
	\section{Introduction}
	\vspace{-0.1cm}
	
Integrating sensing into wireless communication systems has emerged as a prominent application scenario for future sixth-generation (6G) wireless networks \cite{cui2021integrating}. This catalyzes extensive emerging technologies, including autonomous driving, virtual reality, and airspace supervision. With the wireless sensing capability, cellular base stations (BSs) can extract useful environmental and object information from echo signals \cite{liu2022integrated}, and achieve  ultra-resolution and high-accuracy sensing by exploiting the massive antenna deployment. Furthermore, the base station (BS) with multiple antennas in communication networks is able to provide ultra-resolution and high-accuracy sensing. However, the wireless sensing generally requires the line-of-sight (LoS) channel between the BS and target, which is typically blocked by various infrastructures, vehicles, or vegetation, thus seriously limiting the sensing performance \cite{huang2019reconfigurable}.
	
The development of passive intelligent reflecting surfaces (IRSs) provides a viable solution to bypassing the blockage. This is achieved by constructing virtual LoS paths via reflecting the incident signals with properly controlled phases and/or amplitudes \cite{wu2021intelligent}. In the literature, there have been various existing works aiming to enhance communication or sensing  capabilities by deploying passive IRSs in wireless systems \cite{song2023intelligent,fang2023multiirsenabled,song2024cramer}. Nevertheless, transmit signals generally suffer from significant path loss caused by multi-hop reflections, which forms the bottleneck for further improving the system performance \cite{buzzi2022foundations}. To overcome this limitation, a new {\it active} IRS architecture has been proposed in \cite{zhang2022active}. In contrast to passive IRSs, which only reflect signals without amplification, active IRSs have the capability of amplifying reflecting signals through the integration of reflection-type amplifiers into reflecting elements. Despite the additional power consumption, active IRS can effectively compensate for the severe path loss in an energy-efficient manner \cite{kang2024active}.
	
The merits of {active IRSs} make them excellent enablers for both communication and sensing systems. Several works conducted preliminary studies on the application of active IRSs in integrated sensing and communication (ISAC) systems. For instance, in \cite{salem2022active}, the authors proposed the utilization of an active IRS to enhance the communication secrecy rate, aiming at maximizing the worst radar detection signal-to-noise ratio (SNR). Furthermore, \cite{zhang2022CRAN} investigated an active IRS-aided ISAC system in a cloud radio access network, where an active IRS is adopted to address the blockage issue between the BS and targets/users. The radar beampattern towards the sensing targets was optimized to boost the sensing performance. Similarly, the work \cite{zhu2023joint} also deployed an active IRS to introduce an additional virtual LoS link between the BS and the target. The study focused on the joint design of transmit/receive and reflective beamforming to maximize the radar SNR while ensuring the predefined signal-to-interference-plus-noise ratios (SINRs) for communication users. However, these prior works on active IRS-enabled sensing or ISAC mainly focused on the case with one single IRS. Unfortunately, for sensing tasks such as direction-of-arrival (DoA)-based target localization, a single IRS only provides a piece of sensing information inferred from one observation angle, which is not robust and far from satisfactory for target sensing. Therefore, leveraging multiple IRSs to achieve multi-view sensing and provide information from different observation angles is an effective way to improve the performance and robustness of sensing, which motivates our study in this work.

This paper studies a multi-active-IRS cooperative sensing system, in which multiple active IRSs are deployed at different locations to assist the BS in providing multi-view sensing and overcoming the severe path loss due to multi-hop reflections. Specifically, we first propose a multi-IRS cooperative sensing framework based on time division mode, and then derive the closed-form Cram\'{e}r-Rao bound (CRB) for the estimation of target's DoA with respect to each IRS. To achieve the optimal sensing performance, we minimize the maximum CRB for target DoA estimation among all IRSs by jointly optimizing the transmit beamforming at the BS and the reflective beamforming at the IRSs. To obtain a high-quality solution to the highly non-convex max-CRB minimization problem, we propose an efficient algorithm based on alternating optimization, successive convex approximation (SCA), and semi-definite relaxation (SDR). Finally, numerical results verify the effectiveness of our proposed design and the advantages of active IRS-assisted sensing compared to that with passive IRSs. It is shown that the maximum transmit power budget and the maximum amplification gain at the IRSs both limit the sensing performance, especially when the transmit power budget at the BS becomes large. It is also shown that transmit beamforming at the BS is of greater importance than reflective beamforming at IRSs in minimizing the maximum sensing CRB.


	{\it Notations:} The circularly symmetric complex Gaussian distribution with mean $\bm{\mu}$ and covariance $\mathbf{A}$ are denoted as $\mathcal{CN}(\bm{\mu},\mathbf{A})$. The notations $(\cdot)^{T}$, $(\cdot)^{*}$, $(\cdot)^{H}$, and $\mathrm{tr}(\cdot)$ denote the transpose, conjugate, conjugate-transpose, and trace operators, respectively. $\mathbf{I}_{L}$ stands for the identity matrix of size $L \times L$. $\Re(\cdot)$ and $\Im(\cdot)$ denote the real and imaginary parts of the argument, respectively. $|\cdot|$ and $\mathrm{arg}\left\{\cdot\right\}$ denote the
	absolute value and angle of a complex element, respectively. $\mathbb{ E}(\cdot)$ denotes the expectation operation, $\mathrm{diag}(\mathbf{x})$ denotes a diagonal matrix with the diagonal entries specified by vector $\mathbf{x}$, and $\mathrm{Diag}(\mathbf{X})$ denotes a diagonal matrix with the diagonal entries specified by the diagonal elements in $\mathbf{X}$. $\mathrm{rank}\left(\mathbf{X}\right)$ denotes the rank value of matrix $\mathbf{X}$ and $[\cdot]_{l,p}$ denotes the $(l,p)$-th element of a matrix. $j$ denotes the imaginary unit. $\otimes$ and $\circ$ denote the Kronecker product and Hadamard product operators, respectively.
	\vspace{-0.1cm}
	\section{System Model}
	\vspace{-0.1cm}
	\begin{figure}[H]
		\setlength{\abovecaptionskip}{-0pt}
		\setlength{\belowcaptionskip}{-10pt}
		\centering
		\includegraphics[width=0.38\textwidth]{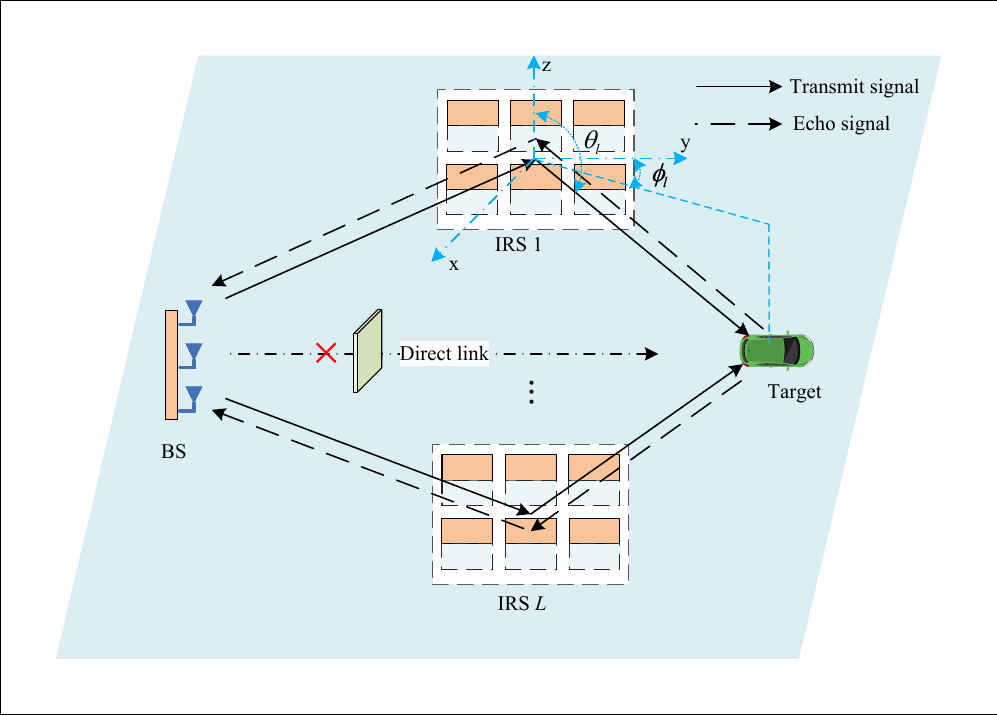}
		\caption{Multi-IRS-assisted cooperative sensing.}\label{fig:SystemModel}
	\end{figure}
	
	We consider a multi-active-IRS cooperative sensing system, where a BS and $L$ active IRSs cooperate to locate one target. The BS is equipped with $M$ uniform linear array (ULA) antennas while each IRS consists of a uniform planar array (UPA) of $N = N_{h}\times N_{v}$ reflecting elements. Let $\mathcal{L} = \{1,\ldots,L\}$ denote the set of IRSs and $\mathcal{N} = \{1,\ldots,N\}$ denote the set of reflecting elements at each IRS in this system. It is assumed that the direct link between the BS and the target is typically obstructed by blockages such as infrastructures, vehicles, and environmental elements. 
	
	We consider the quasi-static channel model, in which the wireless channels remain unchanged over the transmission block of interest. Let $T_{c}$ denote the transmission block of interest or the radar dwell time. $\mathbf{G}_{l} \in \mathbb{C}^{N \times M}$ denotes the channel matrix between the BS and IRS $l$, which can be obtained by the system via proper channel estimation (see, e.g., \cite{zheng2022survey}). We assume that the IRSs are delicately deployed such that the channels between IRSs and the target are LoS. Denote $\theta_{l}$ and $\phi_{l}$ as the vertical and azimuth angles of the target with respect to IRS $l$, respectively. The steering vector of the reflecting elements at IRS $l$ for the DoA $(\theta_{l},\phi_{l})$ is given by
	\begin{align}
			\mathbf{a}_{l} = \mathbf{a}_{v}(\theta_l)\otimes\mathbf{a}_{h}(\theta_l,\phi_l),
		\end{align} 
	where
		\begin{align}
			&\mathbf{a}_{v}(\theta_l)=[1,e^{j\frac{2\pi d_v}{\lambda}\cos(\theta_l)},\cdots,e^{j\frac{2\pi d_v(N_v -1)}{\lambda}\cos(\theta_l)}]^{T},\\
			&\mathbf{a}_{h}(\theta_l,\phi_l)=\nonumber\\
			&[1,e^{j\frac{2\pi d_h}{\lambda}\sin(\theta_l)\cos(\phi_l)},\cdots,e^{j\frac{2\pi d_h(N_h -1)}{\lambda}\sin(\theta_l)\cos(\phi_l)}]^{T},
		\end{align}
		and $\lambda$ denotes the carrier wavelength. In addition, $d_h$ and $d_v$ denote the horizontal and vertical spacing of two neighboring reflection elements at IRSs, respectively. Thus, the round-trip target response matrix of the IRS $l'$-target-IRS $l$ link is given by
		\vspace{-2mm}
		\begin{align}
			\mathbf{E}_{l,l'} = \beta_{l,l'}\mathbf{a}_{l}\mathbf{a}_{l'}^{T},\label{Target_RM_reflect}
		\end{align}
		where $\beta_{l,l'}$ denotes the complex coefficient that accounts for the radar cross-section of the target and the round-trip path loss of the IRS $l'$-target-IRS $l$ link. Furthermore, we define $\beta_{l} \triangleq \beta_{l,l}$ and $\mathbf{E}_{l} \triangleq \mathbf{E}_{l,l} $ for ease of presentation.

	
	
	Next, we consider the design of transmit beamforming at the BS and the reflective beamforming at the IRSs. Let $\bm{\psi}_{l} = [\psi_{l,1},\ldots, \psi_{l,N}]^{T}$ denote the complex reflection coefficients imposed by IRS $l$ and $a_{\text{max}}$ denote the maximum amplitude amplification gain of the elements at IRS. Since each element of active IRSs can not only tune the phase but also amplify the amplitude of the signal, the complex reflective coefficient can be formulated as $\psi_{l,n} = a_{l,n}e^{j\rho_{l,n}}, \forall l\in \mathcal{L}, n\in \mathcal{N}$, where $a_{l,n}$ and $\rho_{l,n}$ are the amplitude and phase of reflective coefficient, respectively \cite{zhang2022active}. The maximum power amplification gain constraints at the active IRSs are expressed as $|\psi_{l,n}|=a_{l,n} \leq a_{\text{max}}, \forall l\in \mathcal{L}, n\in \mathcal{N}$. 
	
	In this work, we consider that the IRSs operate in a time division mode to avoid the inter-IRS interference, where the total radar dwell time $T_{c}$ is divided into $L$ time blocks, each consisting of $T_{L} = \frac{T_{c}}{L}$ time symbols. Under this mode, IRS $l$ is active during the time symbols $\{(l-1)\frac{T_{c}}{L}+1,\ldots,l\frac{T_{c}}{L}\}$ while the others are silent. Let $\mathbf{s}_{l}[t] = \left[{s}_{l,1}[t],\dots,s_{l,M}[t]\right]^{T}$ denote the transmitted signal by the BS at time symbol $t\in \{(l-1)\frac{T_{c}}{L}+1,\ldots,l\frac{T_{c}}{L}\}$. Then, the sample covariance matrix of the transmit signal over the corresponding $T_L$ time symbols is given by $
		\mathbf{R}_{s,l} =\mathbb{E}\left\{ \mathbf{s}_{l}[t]\mathbf{s}_{l}^{H}[t] \right\} \simeq  \frac{1}{T_{L}} \sum\nolimits_{t=(l-1)\frac{T_{c}}{L}+1}^{l\frac{T_{c}}{L}}\mathbf{s}_{l}[t]\mathbf{s}_{l}^{H}[t] \succeq 0$.
	The total sample covariance matrix over the radar dwell time $T_{c}$ needs to satisfy the maximum transmit power constraint $\frac{1}{L}\sum\nolimits_{l \in \mathcal{L}}\mathrm{tr}\left(\mathbf{R}_{s,l}\right) \leq P_{\text{t}}$, where $P_{\text{t}}$ is the maximum transmit power at the BS.
	
	Since the BS normally has more computation capacity, in this paper we consider sensing signals are received and processed at the BS. In contrast, the active IRSs only manipulate signals from the BS or echo signals reflected by the target. In particular, each IRS amplifies twice during one round-trip signal propagation. First, IRS $l$ amplifies the transmit signal $\mathbf{s}_{l}[t]$ from the BS as
	\begin{align}
		{\mathbf{x}_{l,1}}[t] &=  \mathbf{\Psi}_{l}\mathbf{G}_{l}\mathbf{s}_{l}[t]  + \mathbf{\Psi}_{l}{\mathbf{z}_{l,1}}[t],\label{IRS_ref_sig11}
	\end{align}
	where $\mathbf{\Psi}_{l} = \text{diag}(\bm{\psi}_{l})$, and ${\mathbf{z}_{l,1}}[t] \sim \mathcal{CN}(\mathbf{0},\sigma_{\text{r}}^2\mathbf{I}_{N})$ denotes the reflected amplification noise induced by active IRS $l$. Second, when the target reflects the signal back to IRS $l$, IRS $l$ again amplifies the echo signal as
		\begin{align}
		{\mathbf{x}_{l,2}}[t] &= \mathbf{\Psi}_{l}\left(\mathbf{E}_{l}{\mathbf{x}_{l,1}}[t] + {\mathbf{z}_{l,2}}[t]\right)\nonumber\\
		&=  \mathbf{\Psi}_{l}\mathbf{E}_{l}\mathbf{\Psi}_{l}\mathbf{G}_{l}\mathbf{s}_{l}[t] + \mathbf{\Psi}_{l}\mathbf{E}_{l}\mathbf{\Psi}_{l}{\mathbf{z}_{l,1}}[t] + \mathbf{\Psi}_{l}{\mathbf{z}_{l,2}}[t],\label{IRS_ref_sig12}
	\end{align}
where ${\mathbf{z}_{l,2}}[t]\sim \mathcal{CN}(\mathbf{0},\sigma_{\text{r}}^2\mathbf{I}_{N})$ is the reflected amplification noise induced by active IRS $l$. Let $P_{\text{s}}$ denote the maximum transmit power budget at each IRS. Therefore, the power constraint at each active IRS is given by   
	\begin{align}
		C1:\;\;&\mathbb{E}\left\{ \|{\mathbf{x}_{l,1}}[t]\|^2 + \|{\mathbf{x}_{l,2}}[t]\|^2 \right\} = \nonumber\\
		&\mathrm{tr}\left(\mathbf{\Psi}_{l}\mathbf{E}_{l}\mathbf{\Psi}_{l}\mathbf{G}_{l}\mathbf{R}_{s,l}\mathbf{G}_{l}^{H}\mathbf{\Psi}_{l}^{H}\mathbf{E}_{l}^{H}\mathbf{\Psi}_{l}^{H}\right) \nonumber\\
		&+ \sigma_{\text{r}}^2\mathrm{tr}\left(\mathbf{\Psi}_{l}\mathbf{E}_{l}\mathbf{\Psi}_{l}\mathbf{\Psi}_{l}^{H}\mathbf{E}_{l}^{H}\mathbf{\Psi}_{l}^{H}\right)\!\!+\!\! \mathrm{tr}\left( \mathbf{\Psi}_{l}\mathbf{G}_{l}\mathbf{R}_{s,l}\mathbf{G}_{l}^{H}\mathbf{\Psi}_{l}^{H} \right)  \nonumber\\
		&+ 2 \sigma_{\text{r}}^2 \mathrm{tr}\left(\mathbf{\Psi}_{l}\mathbf{\Psi}_{l}^{H}\right) \leq P_{\text{s}}, \quad \forall l\in \mathcal{L}.\label{IRS_power_cons1}
	\end{align}	
	Based on the signal model at the IRSs in \eqref{IRS_ref_sig12}, the received echo signal at the BS from IRS $l$ at time symbol $t\in \{(l-1)\frac{T_{c}}{L}+1,\ldots,l\frac{T_{c}}{L}\}$ is given by
	\begin{align} \label{ReflectedRadar}
		{\mathbf{y}}_{l}[t] &=  \mathbf{G}_{l}^{T}\mathbf{\Psi}_{l}\mathbf{E}_{l}\mathbf{\Psi}_{l}\mathbf{G}_{l}\mathbf{s}_{l}[t]  + \mathbf{G}_{l}^{T}\mathbf{\Psi}_{l}\mathbf{E}_{l}\mathbf{\Psi}_{l}{\mathbf{z}_{l,1}}[t] \nonumber\\
		&+ \mathbf{G}_{l}^{T}\mathbf{\Psi}_{l}{\mathbf{z}_{l,2}}[t] + {\mathbf{z}}[t],
	\end{align}
	where $\mathbf{z}[t]\sim \mathcal{CN}(\mathbf{0},\sigma_{\text{b}}^2\mathbf{I}_{M}) $ denotes the additive white Gaussian noise (AWGN) at the BS.  By concatenating $\mathbf{S}_{l}= [\mathbf{s}_{l}[(l-1)\frac{T_{c}}{L}+1],\ldots,\mathbf{s}_{l}[l\frac{T_{c}}{L}]]$, ${\mathbf{Y}}_{l} = [{\mathbf{y}}_{l}[(l-1)\frac{T_{c}}{L}+1],\ldots,{\mathbf{y}}_{l}[l\frac{T_{c}}{L}]]$, ${\mathbf{Z}}_{l,1} = [{\mathbf{z}}_{l,1}[(l-1)\frac{T_{c}}{L}+1],\ldots,{\mathbf{z}}_{l,1}[l\frac{T_{c}}{L}]]$, and ${\mathbf{Z}}_{l} = [{\mathbf{z}}[(l-1)\frac{T_{c}}{L}+1],\ldots,{\mathbf{z}}[l\frac{T_{c}}{L}]]$, we have 
	{\small
	\begin{align}
		\!\!{\mathbf{Y}}_{l} &\!=\!  \mathbf{G}_{l}^{T}\mathbf{\Psi}_{l}\mathbf{E}_{l}\mathbf{\Psi}_{l}\mathbf{G}_{l}\mathbf{S}_{l}  \!+\! \mathbf{G}_{l}^{T}\mathbf{\Psi}_{l}\mathbf{E}_{l}\mathbf{\Psi}_{l}{\mathbf{Z}_{l,1}} \!\!+\! \mathbf{G}_{l}^{T}\mathbf{\Psi}_{l}{\mathbf{Z}_{l,2}} \!+\! {\mathbf{Z}}_{l}.\label{EchoSigAtBS1}
	\end{align}}
	 Accordingly, based on the received echo signal ${\mathbf{Y}}_{l}$ in \eqref{EchoSigAtBS1},  the BS needs to estimate the DoAs $\{\theta_{l},\phi_{l}\}$ and the complex coefficient $\{\beta_{l}\}$ in the complete target response matrix $\mathbf{E}_{l}$ as unknown parameters. Then, based on the estimated DoAs of the target with respect to all IRSs $\{\theta_{l},\phi_{l}\}, l \in \mathcal{L}$, the Stansfield method can be utilized to infer the coordinates of the target \cite{gavish1992performance}.\footnote{The design of DoA estimation and target localization algorithms is beyond the scope of this paper's work, and we defer them for future research.}
	 \vspace{-0.1cm} 
\section{Estimation CRB Derivation for Active IRS}	
In this section, we derive the CRB for target localization. Specifically, the BS estimates the DoA of the target with regard to the IRSs and then uses the multiple estimates of DoAs to infer the 3D position of the target. Let $\bar{\bm{\xi}}_{l} = [\theta_{l}, \phi_{l}, {\bm{\beta}_{l}^{T}}]^{T}$ denote the vector of unknown parameters to be estimated with respect to IRS $l$, where ${\bm{\beta}_{l}} = [\Re(\beta_{l}),\Im({\beta_{l}})]^{T}$. By vectorizing the received echo signal $\mathbf{Y}_{l}$, we have
\begin{align}
	{\mathbf{y}}_{l} & = \mathrm{vec}({\mathbf{Y}}_{l})= {\bm{\eta}}_{l} +{\mathbf{w}}_{l},\label{vec_y_l_bs}
\end{align}
where
\begin{align}
	{\bm{\eta}}_{l} &= \left[(\mathbf{G}_{l}^{T}\mathbf{\Psi}_{l}{\mathbf{E}}_{l}\mathbf{\Psi}_{l}\mathbf{G}_{l}\mathbf{s}_{l}[\tfrac{(l-1)T_{c}}{L}+1])^{T}, \right.\nonumber\\
	&\left.\cdots,(\mathbf{G}_{l}^{T}\mathbf{\Psi}_{l}{\mathbf{E}}_{l}\mathbf{\Psi}_{l}\mathbf{G}_{l}\mathbf{s}_{l}[l\tfrac{T_{c}}{L}])^{T} \right]^{T},\label{vec_sig_bs}\\
	{\mathbf{w}}_{l} &= \left[(\mathbf{G}_{l}^{T}\mathbf{\Psi}_{l}{\mathbf{z}_{l,2}}[\tfrac{(l-1)T_{c}}{L}+1])^{T} + ({\mathbf{z}}_{l}[\tfrac{(l-1)T_{c}}{L}+1])^{T}, \right.\nonumber\\
	&\left.\cdots,(\mathbf{G}_{l}^{T}\mathbf{\Psi}_{l}{\mathbf{z}_{l,2}}[l\tfrac{T_{c}}{l}])^{T} + ({\mathbf{z}}_{l}[l\tfrac{T_{c}}{l}])^{T}\right].\label{vec_noise_bs}
\end{align}
Note that in \eqref{vec_noise_bs}, we neglect the effect of the noise term ${\mathbf{z}_{l,1}}$ due to the triple-reflection path loss from the IRS-target-IRS-BS link. Based on \eqref{vec_sig_bs} and \eqref{vec_noise_bs}, the mean and covariance matrix of ${\mathbf{y}}_{l}$ are obtained as ${\bm{\eta}}_{l}$ and 
\begin{align}
	\mathbf{R}_{{\mathbf{y}}_{l}} = \mathbf{I}_{\frac{T_{c}}{L}} \otimes \mathbf{R}_{{\mathbf{w}}_{l}},\label{R_y_bar_bs}
\end{align}
respectively, where $\mathbf{R}_{{\mathbf{w}}_{l}} =\sigma_{\text{r}}^{2}\mathbf{G}_{l}^{T}\mathbf{\Psi}_{l}\mathbf{\Psi}_{l}^{H}\mathbf{G}_{l}^{*}+\sigma_{\text{b}}^{2}\mathbf{I}_{{M}}$. According to the definition of CRB, the CRB for estimating parameter vector $\bar{\bm{\xi}}_{l}$ is given by $\mathrm{CRB}_{\bar{\bm{\xi}}_{l}}(\mathbf{R}_{s,l},\mathbf{\Psi}_{l})=\mathrm{tr}(\mathbf{F}_{l}^{-1})$, where ${\mathbf{F}}_{l} \in \mathbb{R}^{4\times4}$ denotes the Fisher information matrix (FIM) with respect to $\bar{\bm{\xi}}_{l}$. According to the estimation theory, the $(p,q)$-th element of ${\mathbf{F}}_{l}$ is given by \cite{kay1993fundamentals}
{\small
\begin{align}
	&[{\mathbf{F}}_{l}]_{p,q} = \nonumber\\
	&\mathrm{tr}\left(\mathbf{R}_{{\mathbf{y}}_{l}}^{-1}\frac{\partial \mathbf{R}_{{\mathbf{y}}_{l}}}{\partial [\bar{\bm{\xi}}_{l}]_{p}}\mathbf{R}_{{\mathbf{y}}_{l}}^{-1}\frac{\partial \mathbf{R}_{{\mathbf{y}}_{l}}}{\partial [\bar{\bm{\xi}}_{l}]_{q}}\right)\!\! +\!\! 2\Re\left(\frac{\partial {\bm{\eta}}_{l}^{H}}{\partial [\bar{\bm{\xi}}_{l}]_{p} } \mathbf{R}_{{\mathbf{y}}_{l}}^{-1} \frac{\partial {\bm{\eta}}_{l}}{\partial [\bar{\bm{\xi}}_{l}]_{q} }\right). \label{FIM_def_bs}
\end{align}}

Based on \eqref{FIM_def_bs}, we have the following proposition.
\begin{prop}\label{prop_FIM_bs}
	We define the derivatives of ${{\mathbf{a}}}$ with respect to $\theta_{l}$ and $\phi_{l}$ as $\dot{{\mathbf{a}}}_{\theta_{l}}$ and $\dot{{\mathbf{a}}}_{\phi_{l}}$, respectively. Then, the FIM for estimating ${\mathbf{F}}_{l}$ is given by 
	\begin{align}
		{\mathbf{F}}_{l} = \left[\begin{array}{ccc}
			{{F}}_{\theta_{l},\theta_l}&{{F}}_{\theta_{l},\phi_{l}}&{\mathbf{F}}_{\theta_{l},\bm{\beta}_{l}}\\
			{{F}}_{\theta_{l},\phi_{l}}^{T}&{{F}}_{\phi_{l},\phi_{l}}&{\mathbf{F}}_{\phi_{l},\bm{\beta}_{l}}\\
			{\mathbf{F}}_{\theta_{l},\bm{\beta}_{l}}^{T}& {\mathbf{F}}_{\phi_{l},\bm{\beta}_{l}}^{T}&{\mathbf{F}}_{{\bm{\beta}_{l}},\bm{\beta}_{l}}
		\end{array}\right],\label{FIM_bs}
\end{align}
	where 
\begin{subequations}
	\begin{align}
	{{F}}_{\theta_{l},\theta_l} &=\frac{{{2 T_c}}}{L}{\left| {{\beta _l}} \right|^2}\mathrm{tr}\left(  \mathbf{C}_{\theta_{l},l}^H  {\mathbf{R}_{{\mathbf{w}}_{l}}^{ - 1}} \mathbf{C}_{\theta_{l},l} {\mathbf{R}_{s,l}} \right),\label{F_theta_theta}\\
   {{F}}_{\phi_{l},\phi_{l}} &=  \frac{{{2 T_c}}}{L} {\left| {{\beta _l}} \right|^2}\mathrm{tr}\left(  \mathbf{C}_{\phi_{l},l}^H  {\mathbf{R}_{{\mathbf{w}}_{l}}^{ - 1}}  \mathbf{C}_{\phi_{l},l} {\mathbf{R}_{s,l}} \right),\label{F_phi_phi}\\
   {\mathbf{F}}_{{\bm{\beta}_{l}},\bm{\beta}_{l}} &=  \frac{{{2 T_c}}}{L}\mathrm{tr}\left(  \mathbf{H}_{l}^H   {\mathbf{R}_{{\mathbf{w}}_{l}}^{ - 1}} \mathbf{H}_{l} {\mathbf{R}_{s,l}} \right) \mathbf{I}_{2},\label{F_beta_beta}\\ 
   {{F}}_{\theta_{l},\phi_{l}} &= \frac{{{2 T_c}}}{L} {\left| {{\beta _l}} \right|^2}\mathrm{tr}\left(  \mathbf{C}_{\theta_{l},l}^H  {\mathbf{R}_{{\mathbf{w}}_{l}}^{ - 1}} \mathbf{C}_{\phi_{l},l}{\mathbf{R}_{s,l}} \right),\label{F_theta_phi}\\
   {\mathbf{F}}_{\theta_{l},\bm{\beta}_{l}} &= \frac{{{2 T_c}}}{L}\Re \left( \beta_{l}\mathrm{tr}\left( \mathbf{C}_{\theta_{l},l}^H  {\mathbf{R}_{{\mathbf{w}}_{l}}^{ - 1}}\mathbf{H}_{l} {\mathbf{R}_{s,l}} \right)[1, j]\right),\label{F_theta_beta}\\
   {\mathbf{F}}_{\phi_{l},\bm{\beta}_{l}} &= \frac{{{2 T_c}}}{L}\Re \left( \beta_{l}\mathrm{tr}\left(\mathbf{C}_{\phi_{l},l}^H   {\mathbf{R}_{{\mathbf{w}}_{l}}^{ - 1}} \mathbf{H}_{l} {\mathbf{R}_{s,l}} \right)[1, j]\right),\label{F_phi_beta}
\end{align}
\end{subequations}
with $\mathbf{C}_{\theta_{l},l} = \mathbf{G}_{l}^{T}\mathbf{\Psi}_{l} \left({\dot{{\mathbf{a}}}_{\theta_{l}} {\mathbf{a}} _l^T + {{ {\mathbf{a}} }_l}{\dot{{\mathbf{a}}}_{\theta_{l}}^T}}\right) {{\mathbf{\Psi }}_l}{{\mathbf{G}}_l}$, $\mathbf{C}_{\phi_{l},l} = \mathbf{G}_{l}^{T}\mathbf{\Psi}_{l} \left({\dot{{\mathbf{a}}}_{\phi_{l}} {\mathbf{a}} _l^T + {{ {\mathbf{a}} }_l}{\dot{{\mathbf{a}}}_{\phi_{l}}^T}}\right) {{\mathbf{\Psi }}_l}{{\mathbf{G}}_l}$, and $\mathbf{H}_{l} =\mathbf{G}_{l}^{T}\mathbf{\Psi}_{l} {{\mathbf{a}}}_{l} {\mathbf{a}} _l^T  {{\mathbf{\Psi }}_l}{{\mathbf{G}}_l}$.
\end{prop}
\begin{IEEEproof}
	See Appendix \ref{prop_FIM1_proof}.	
\end{IEEEproof}
\vspace{-0.2cm} 
\section{Joint Transmit and Reflective Beamforming for Target Estimation}
In this section, we jointly design the transmit beamforming $\{\mathbf{R}_{s,l}\}$ at the BS and the reflective beamforming $\{\mathbf{\Psi}_{l}\}$ at the IRSs to improve the performance of target estimation. Specifically, we first use the derived closed-form FIM in \eqref{FIM_bs} to formulate the optimization problem. Then, we propose an efficient algorithm to find a high-quality solution based on alternating optimization.

Our aim is to minimize the maximum CRBs among $\mathrm{CRB}_{\bar{\bm{\xi}}_{l}}(\mathbf{R}_{s,l},\mathbf{\Psi}_{l})$, subject to the maximum transmit power constraint at the BS, the maximum transmit power constraints at the IRSs, and the maximum power amplification gain constraints at the IRSs. Consequently, the optimization problem is formulated as 
\vspace{-2mm}
\begin{subequations}
	\begin{eqnarray}
			\!\!\!(\text{P1}):\!\!\!&{\mathop {\min}\limits_{\{\boldsymbol{\Psi}_{l},\mathbf{R}_{s,l}\} } } \!\!\!\!&\mathop{\max} \limits_{l \in \mathcal{L}}  \quad \mathrm{CRB}_{\bar{\bm{\xi}}_{l}}(\mathbf{R}_{s,l},\mathbf{\Psi}_{l}) \nonumber\\
		&\text{s.t.} \!\!\!\!\!\!\!& C1, \label{P1_I_cons1}\\
		&& \frac{1}{L}\sum\nolimits_{l \in \mathcal{L}}\mathrm{tr}\left( \mathbf{R}_{s,l}\right) \leq P_{\text{t}}, \label{P1_I_cons2}\\	
		&&\mathbf{R}_{s,l} \succeq \boldsymbol{0}, \label{P1_I_cons3}\\
		&&|[\mathbf{\Psi}_{l}]_{n,n}| \leq a_{\text{max}}, \forall l\in \mathcal{L}, n \in \mathcal{N}. \label{P1_I_cons4}
	\end{eqnarray}
\end{subequations}
In problem (P1), \eqref{P1_I_cons1}-\eqref{P1_I_cons4} are the transmit power constraints at the IRSs, the transmit power constraint at the BS, the semi-definite constraint regarding the sample covariance matrix of the transmit signal, and the maximum power amplification gain constraints at the IRSs, respectively. Note that problem (P1) is non-convex due to the non-convexity of the objective function and the constraint in \eqref{P1_I_cons1}. To address this issue, we adopt the alternating optimization approach, wherein the transmit signal covariance $\{\mathbf{R}_{s,l}\}$ at the BS and the reflection coefficients $\{\mathbf{\Psi}_{l}\}$ at the IRSs are optimized alternately. 
\subsection{Optimal Transmit Beamforming Design}
Under given Reflective Beamforming $\{\mathbf{\Psi}_{l}\}$, the optimization problem is rewritten as 
	\begin{align}
		     \begin{array}{*{20}{lcl}}
			(\text{P2}):&{\mathop {\min}\limits_{\{\mathbf{R}_{s,l}\} } } &\mathop{\max} \limits_{l \in \mathcal{L}}  \quad \mathrm{tr}(\mathbf{F}_{l}^{-1}) &\text{s.t.} & \eqref{P1_I_cons1}-\eqref{P1_I_cons3}.\nonumber
		\end{array}
	\end{align}
Problem (P2) is also difficult to solve due to the fact that the objective function cannot be expressed in an analytical form. To solve problem (P2), we first introduce an auxiliary variable $\kappa_{i}$ to transform it into the following equivalent problem:
\begin{subequations}
	\begin{eqnarray}
			(\text{P2.1}):&{\mathop {\min}\limits_{ \{\mathbf{R}_{s,l}\},\kappa  } } &  \kappa \nonumber\\
		&\text{s.t.} & \mathrm{tr}(\mathbf{F}_{l}^{-1}) \leq \kappa, \forall l\in \mathcal{L} \label{P2_1_cons4}, \\
		&& \eqref{P1_I_cons1}-\eqref{P1_I_cons3}.\nonumber
	\end{eqnarray}
\end{subequations}
From the expression of FIM in \eqref{FIM_bs}, we note that $\mathbf{F}_{l}$ is a linear function to $\mathbf{R}_{s,l}$ and thus $\mathrm{tr}(\mathbf{F}_{l}^{-1})$ is a convex function to $\mathbf{R}_{s,l}$. Therefore, the constraints in \eqref{P2_1_cons4} are convex. As a result, problem (P2.1) is a convex semi-definite program (SDP) problem, which can be optimally solved by existing solvers like CVX \cite{grant2014cvx}.  
\subsection{Reflective Beamforming Design}
Under given transmit beamforming $\{\mathbf{R}_{s,l}\}$, the optimization problem is rewritten as
	\begin{align}
			\begin{array}{*{20}{lcl}}
				(\text{P3}):&{\mathop {\min}\limits_{\{\boldsymbol{\Psi}_{l}\} } } &\mathop{\max} \limits_{l \in \mathcal{L}}  \quad \mathrm{tr}(\mathbf{F}_{l}^{-1}) 	&\text{s.t.} & \eqref{P1_I_cons1}, \eqref{P1_I_cons4}.\nonumber
			\end{array}
	\end{align}
Note that the FIM $\mathbf{F}_{l}$ depends only on the reflective beamforming $\boldsymbol{\Psi}_{l}$ at the IRS $l$. Thus, problem (P3) can be equivalently decomposed into $L$ subproblems each given by
	\begin{align}
			\begin{array}{*{20}{lcl}}
				(\text{P3.$l$.1}):&{\mathop {\min}\limits_{\boldsymbol{\Psi}_{l} } } & \mathrm{tr}(\mathbf{F}_{l}^{-1})&\text{s.t.} & \eqref{P1_I_cons1}, \eqref{P1_I_cons4}.\nonumber
			\end{array}
	\end{align}
	
 This problem is highly non-convex because the transmit power constraints at IRSs in \eqref{P1_I_cons1} and the elements in FIM are non-convex functions with respect to $\{\boldsymbol{\Psi}_{l}\}$. By introducing auxiliary variables $\{\kappa_i\}_{i=1}^{4}$, problem (\text{P3.$l$.1}) is equivalent to 
 \begin{subequations}
	\begin{eqnarray}
				\!\!\!\!\!(\text{P3.$l$.2}):&\!\!\!\!\!\!\!{\mathop {\min}\limits_{\boldsymbol{\Psi}_{l},\{\kappa_i\}_{i=1}^{4}  } } & \sum\nolimits_{i=1}^{4} \kappa_i \nonumber\\
			&\text{s.t.} & \left[\begin{array}{cc}
				\mathbf{F}_{l} & \mathbf{e}_{i} \\
				 \mathbf{e}_{i}^{T}  & \kappa_i 
			\end{array}\right] \succeq 0, i = 1, \ldots, 4, \label{Schur_Com}\\
			&&\eqref{P1_I_cons1}, \eqref{P1_I_cons4},\nonumber
	\end{eqnarray}
\end{subequations}
where $\mathbf{e}_{i}$ denotes the $i$-th column of the identity matrix $\mathbf{I}_{4}$, and the constraint in \eqref{Schur_Com} is derived by using the Schur complement. Note that problem (P3.$l$.2) is still non-convex due to the constraints in \eqref{P1_I_cons1} and \eqref{Schur_Com}. To handle it, we resort to SDR and SCA techniques to transform these constraints into a convex form.

First, by defining $\mathbf{\Theta}_{l} = \bm{\psi}_{l}\bm{\psi}_{l}^{H}$, where $\mathrm{rank}(\mathbf{\Theta}_{l} ) = 1$, we approximate the objective function $\mathrm{tr}(\mathbf{F}_{l}^{-1})$ according to Lemma \ref{FIM_appr_lem}. 
\begin{lem}\label{FIM_appr_lem}
With a given local point $\mathbf{\Theta}_{l}^{(i)}$, the FIM in \eqref{FIM_bs} is approximated as 
\begin{align}
	\hat{{\mathbf{F}}}_{l} = \left[\begin{array}{ccc}
		\hat{{{F}}}_{\theta_{l},\theta_l}&\hat{{{F}}}_{\theta_{l},\phi_{l}}&\hat{{\mathbf{F}}}_{\theta_{l},\bm{\beta}_{l}}\\
		\hat{{{F}}}_{\theta_{l},\phi_{l}}^{T}&\hat{{{F}}}_{\phi_{l},\phi_{l}}&\hat{{\mathbf{F}}}_{\phi_{l},\bm{\beta}_{l}}\\
		\hat{{\mathbf{F}}}_{\theta_{l},\bm{\beta}_{l}}^{T}& \hat{{\mathbf{F}}}_{\phi_{l},\bm{\beta}_{l}}^{T}&\hat{{\mathbf{F}}}_{{\bm{\beta}_{l}},\bm{\beta}_{l}}
	\end{array}\right],\label{FIM_bs_trans}
\end{align}
in which 
{\small
\begin{subequations}
	\begin{align}
	\hat{{F}}_{\varrho_{1}} &= \frac{{{2 T_c}}}{L}{\left| {{\beta _l}} \right|^2} \mathrm{tr}\left(\nabla_{\varrho_{1}}^{T}(\mathbf{\Theta}_{l}^{(i)}) \left(\mathbf{\Theta}_{l}-\mathbf{\Theta}_{l}^{(i)}\right)  \right) \nonumber\\
	&+ \frac{{{2 T_c}}}{L}{\left| {{\beta _l}} \right|^2} Q_{\varrho_{1}}(\mathbf{\Theta}_{l}^{(i)}),\label{hat_F_theta_theta}\\
   \hat{\mathbf{F}}_{\varrho_{2}} &= \frac{{{2 T_c}}}{L}\Re \left( \beta_{l} \left( \mathrm{tr} \left(\nabla_{\varrho_{2}}^{T}(\mathbf{\Theta}_{l}^{(i)}) \left(\mathbf{\Theta}_{l} - \mathbf{\Theta}_{l}^{(i)}\right) \right) \right.\right. \nonumber\\
   &\left.\left. + Q_{\varrho_{2}}(\mathbf{\Theta}_{l}^{(i)}) \right) [1, j]\right),\label{hat_F_phi_beta}\\
   \hat{\mathbf{F}}_{{\bm{\beta}_{l}},\bm{\beta}_{l}} &=  \frac{{{2 T_c}}}{L} \left( \mathrm{tr}\left(\nabla_{{\bm{\beta}_{l}},\bm{\beta}_{l}}^{T}(\mathbf{\Theta}_{l}^{(i)})\left(\mathbf{\Theta}_{l} - \mathbf{\Theta}_{l}^{{i}}\right)  \right) \right. \nonumber\\
   &\left.+ Q_{{\bm{\beta}_{l}},\bm{\beta}_{l}}(\mathbf{\Theta}_{l}^{(i)}) \right) \mathbf{I}_{2},\label{hat_F_beta_beta}
\end{align}
\end{subequations}}where $\varrho_{1} \in \{(\theta_{l},\theta_l),(\phi_{l},\phi_{l}),(\theta_{l},\phi_{l})\}$ and $\varrho_{2} \in \{(\theta_{l},\bm{\beta}_{l}),(\phi_{l},\bm{\beta}_{l})\}$.
\end{lem}
\begin{IEEEproof}
	See Appendix \ref{FIM_appr_proof}.
\end{IEEEproof}

Then, we transform the constraint in \eqref{P1_I_cons1} into a convex form. The four terms in the left-hand-side of \eqref{P1_I_cons1} are equivalent to 
\begin{subequations} 
	\begin{align}
	&\mathrm{tr}\left(\mathbf{\Psi}_{l}\mathbf{E}_{l}\mathbf{\Psi}_{l}\mathbf{G}_{l}\mathbf{R}_{s,l}\mathbf{G}_{l}^{H}\mathbf{\Psi}_{l}^{H}\mathbf{E}_{l}^{H}\mathbf{\Psi}_{l}^{H}\right)\nonumber\\
	&= \left| {{\beta _l}} \right|^2 \mathrm{tr}\left(\mathbf{\Psi}_{l} \mathbf{a}_{l}\mathbf{a}_{l}^{T} \mathbf{\Psi}_{l}\mathbf{G}_{l}\mathbf{R}_{s,l}\mathbf{G}_{l}^{H}\mathbf{\Psi}_{l}^{H}\mathbf{a}_{l}^{*} \mathbf{a}_{l}^{H} \mathbf{\Psi}_{l}^{H}\right) \nonumber\\
	&= \left| {{\beta _l}} \right|^2 \mathrm{tr}\left(\mathbf{A}_{l}^{H}\mathbf{A}_{l}\mathbf{\Theta}_{l}\right) \mathrm{tr}\left(\mathbf{R}_{1}\mathbf{\Theta}_{l}^{T} \right), \label{P1_I_cons1_1}\\
	& \mathrm{tr}\left(\mathbf{\Psi}_{l}\mathbf{E}_{l}\mathbf{\Psi}_{l}\mathbf{\Psi}_{l}^{H}\mathbf{E}_{l}^{H}\mathbf{\Psi}_{l}^{H}\right) \nonumber\\
	&= \left| {{\beta _l}} \right|^2 \mathrm{tr}\left(\mathbf{\Psi}_{l}\mathbf{a}_{l}\mathbf{a}_{l}^{T}\mathbf{\Psi}_{l}\mathbf{\Psi}_{l}^{H}\mathbf{a}_{l}^{*}\mathbf{a}_{l}^{H}\mathbf{\Psi}_{l}^{H}\right) \nonumber\\
	&= \left| {{\beta _l}} \right|^2 \left(\mathrm{tr}\left(\mathbf{A}_{l}^{H}\mathbf{A}_{l}\mathbf{\Theta}_{l}\right)\right)^{2}, \label{P1_I_cons1_2}\\
	& \mathrm{tr}\left( \mathbf{\Psi}_{l}\mathbf{G}_{l}\mathbf{R}_{s,l}\mathbf{G}_{l}^{H}\mathbf{\Psi}_{l}^{H} \right) = \mathrm{tr}\left(\mathbf{G}_{l}\mathbf{R}_{s,l}\mathbf{G}_{l}^{H}\mathrm{Diag}\left(\mathbf{\Theta}_{l}\right) \right), 	 \label{P1_I_cons1_3}\\  
	&\mathrm{tr}\left(\mathbf{\Psi}_{l}\mathbf{\Psi}_{l}^{H}\right) = \mathrm{tr}\left(\mathbf{\Theta}_{l}\right), \label{P1_I_cons1_4}
\end{align}
\end{subequations}
respectively, where $\mathbf{A}_{l} = \mathrm{diag}(\mathbf{a}_{l})$ and $\mathbf{R}_{1} = \mathbf{A}_{l}{{\mathbf{G}}_l} {\mathbf{R}_{s,l}} \mathbf{G}_{l}^{H} \mathbf{A}_{l}^{H}$. Note that \eqref{P1_I_cons1_1} and \eqref{P1_I_cons1_2} are quadratic functions with respect to $\mathbf{\Theta}_{l}$, and we approximate them using their first-order Taylor expansions. The derivatives of the trace terms in \eqref{P1_I_cons1_1} and \eqref{P1_I_cons1_2} with respect to $\mathbf{\Theta}_{l}$ are given by
{\small
\begin{subequations} 
\begin{align}
	\nabla_{\text{c},1}(\mathbf{\Theta}_{l}) & = \frac{\partial}{\partial \mathbf{\Theta}_{l}}   \mathrm{tr}\left(\mathbf{A}_{l}^{H}\mathbf{A}_{l}\mathbf{\Theta}_{l}\right) \mathrm{tr}\left(\mathbf{R}_{1}\mathbf{\Theta}_{l}^{T} \right) \nonumber\\
	&= \mathrm{tr}\left(\mathbf{R}_{1}\mathbf{\Theta}_{l}^{T} \right)\mathbf{A}_{l}^{T}\mathbf{A}_{l}^{*} + \mathrm{tr}\left(\mathbf{A}_{l}^{H}\mathbf{A}_{l}\mathbf{\Theta}_{l}\right)\mathbf{R}_{1}, \label{P1_I_cons1_1_der}\\
	\nabla_{\text{c},2}(\mathbf{\Theta}_{l}) & = \frac{\partial}{\partial \mathbf{\Theta}_{l}}  \left(\mathrm{tr}\left(\mathbf{A}_{l}^{H}\mathbf{A}_{l}\mathbf{\Theta}_{l}\right)\right)^{2} \nonumber\\
	&= 2\mathrm{tr}\left(\mathbf{A}_{l}^{H}\mathbf{A}_{l}\mathbf{\Theta}_{l}\right)\mathbf{A}_{l}^{T}\mathbf{A}_{l}^{*}. \label{P1_I_cons1_2_der}
\end{align}
\end{subequations}}Based on \eqref{P1_I_cons1_3} and \eqref{P1_I_cons1_4}, and the derivatives in \eqref{P1_I_cons1_1_der} and \eqref{P1_I_cons1_2_der}, the constraint in \eqref{P1_I_cons1} is approximated as
{\small
\begin{align}
	&\!\! \!\! \left| {{\beta _l}} \right|^2  \mathrm{tr}\left(\nabla_{\text{c},1}^{T}(\mathbf{\Theta}_{l}^{i})\left(\mathbf{\Theta}_{l}-\mathbf{\Theta}_{l}^{(i)}\right)\right)  \nonumber\\
	&\!\! \!\! + \left| {{\beta _l}} \right|^2 \mathrm{tr}\left(\mathbf{A}_{l}^{H}\mathbf{A}_{l}\mathbf{\Theta}_{l}^{(i)}\right) \mathrm{tr}\left(\mathbf{R}_{1}\left(\mathbf{\Theta}_{l}^{(i)}\right)^{T} \right)  \nonumber\\
	&\!\! \!\!  + \sigma_{\text{r}}^2 \left| {{\beta _l}} \right|^2  \mathrm{tr}\left(\nabla_{\text{c},2}^{T}(\mathbf{\Theta}_{l}^{i})\left(\mathbf{\Theta}_{l}-\mathbf{\Theta}_{l}^{(i)}\right)\right) \nonumber\\
	&\!\! \!\! + \sigma_{\text{r}}^2 \left| {{\beta _l}} \right|^2 \left|\mathrm{tr}\left(\mathbf{A}_{l}^{H}\mathbf{A}_{l}\mathbf{\Theta}_{l}^{(i)}\right)\right|^{2} \nonumber\\
	&\!\! \!\! +\! \mathrm{tr}\left(\mathbf{G}_{l}\mathbf{R}_{s,l}\mathbf{G}_{l}^{H}\mathrm{Diag}\left(\mathbf{\Theta}_{l}\right) \right) \!+\! 2 \sigma_{\text{r}}^2 \mathrm{tr}\left(\mathbf{\Theta}_{l}\right) \leq P_{\text{s}}, \forall l\in \mathcal{L}. \label{P1_I_cons1_trans}
\end{align}}By adopting \eqref{FIM_bs_trans} and \eqref{P1_I_cons1_trans}, in the $i$-th SCA iteration, problem (P3.$l$.2) is transformed into problem (P3.$l$.3) with 
\begin{subequations}
	\begin{eqnarray}
				\!\!\!\!\!(\text{P3.$l$.3}):\!\!\!\!\!\!\!\!&{\mathop {\min}\limits_{\mathbf{\Theta}_{l},\{\kappa_i\}_{i=1}^{4}  } } & \sum\nolimits_{i=1}^{4} \kappa_i \nonumber\\
			&\text{s.t.} & \left[\begin{array}{cc}
				\hat{\mathbf{F}}_{l} & \mathbf{e}_{i} \\
				 \mathbf{e}_{i}^{T}  & \kappa_i 
			\end{array}\right] \succeq 0, i = 1, \ldots, 4, \label{Schur_Com_trans}\\
			&& [\mathbf{\Theta}_{l}]_{n,n} \leq a_{\text{max}}^{2}, \label{amplify_cons}\\
			&& \mathrm{rank}(\mathbf{\Theta}_{l} ) = 1, \label{rank1_cons}\\
			&&\eqref{P1_I_cons1_trans}.\nonumber
	\end{eqnarray}
\end{subequations}
Furthermore, we drop the rank-one constraint in \eqref{rank1_cons} and accordingly obtain the relaxed version of (P3.$l$.3) as (SDR3.$l$.3). Note that problem (SDR3.$l$.3) is a convex problem which can be efficiently solved by CVX. Let $\mathbf{\Theta}_{l}^{\star}$ denote the optimal solution to problem (SDR3.$l$.3), which generally may not meet the rank-one condition. Therefore, we implement Gaussian Randomization to find an efficient rank-one solution to (P3.$l$.3) and (P3.$l$.2) based on the obtained $\{\mathbf{\Theta}_{l}^{\star}\}$ \cite{luo2010semidefinite}. In particular, we generate a number of random vectors ${\boldsymbol{r}}_{l} \sim \mathcal{CN}\left(\boldsymbol{0},\boldsymbol{I}_N\right)$, and then construct a number of rank-one solutions as $\boldsymbol{\psi}_{l} = \left(\boldsymbol{\Theta}_{l}^{\star}\right)^{\frac{1}{2}}\boldsymbol{r}_{l}$. Next, we verify whether the maximum amplitude of the elements in $\boldsymbol{\psi}_{l}$ exceeds the maximum amplification gain $a_{\text{max}}$. If so, we normalize $\boldsymbol{\psi}_{l} = a_{\text{max}}\frac{\boldsymbol{\psi}_{l}}{\max(|\boldsymbol{\psi}_{l}|)}$, where $\max(|\boldsymbol{\psi}_{l}|)$ represents the maximum amplitude value of the elements in $\boldsymbol{\psi}_{l}$. Finally, we seek the optimal solution of $\boldsymbol{\psi}_{l}$ that minimizes $\mathrm{CRB}_{\bar{\bm{\xi}}_{l}}(\mathbf{R}_{s,l},\mathbf{\Psi}_{l})$ while satisfying the constraints in \eqref{P1_I_cons1} among all randomly generated $\boldsymbol{\psi}_{l}$'s. 

Through alternately solving problems (P2) and (P3), a high-quality solution to problem (P1) is obtained. Note that problem (P2) is optimally solved, while solving (P3) leads to a non-increasing sequence of max-CRB values with a sufficiently large number of Gaussian randomizations. Consequently, the alternating optimization-based algorithm guarantees convergence by generating a monotonically decreasing sequence of max-CRB values throughout the iterations. 

After implementing the designed transmit and reflective beamforming in the considered system, the received echo signal in \eqref{EchoSigAtBS1} is leveraged to estimate the DoA $\{\theta_{l},\phi_{l}\}$ by using sophisticated estimation methods, such as the multiple signal classification and rotational invariance techniques \cite{shao2022target}.

\vspace{-0.2cm}
\section{Numerical Results}	
This section provides numerical results to validate the effectiveness of our proposed design. In the simulation, we adopt the Rician fading channel model with the K-factor being $5$ dB for channels between the BS and IRSs. The channels between IRSs and the target are assumed to be LoS channels. Additionally, we set the noise power as $\sigma_{\text{r}}^{2}= \sigma_{\text{b}}^{2} = -80$ dBm, and the radar dwell time as $T_c = 100$ time symbols.
In particular, we consider a scenario with one BS, two active IRSs, and one target. The BS and two IRSs are located at $(0,0,0)$ meters~(m),  $(-5,10,0)$ m, and $(-5,20,0)$ m, respectively. The target is located at $(5,15,0)$ m. To better illustrate the superiority of our proposal, we adopt the following two benchmarks for comparison.

\textbf{Transmit beamforming (BF) only}: The IRSs implement random reflection coefficients. Accordingly, we only optimize the transmit beamforming at the BS by solving problem (P2) in Section IV-A. 
		
\textbf{Reflective BF only}: The BS adopts the isotropic transmission by setting $\mathbf{R}_{s,l}=\frac{P_{\text{t}}}{M}\mathbf{I}_{M}, \forall l\in \mathcal{L}$. Then, we optimize the reflective beamforming at all IRSs by solving problem (P3) in Section IV-B.

\begin{figure}[tbp]
	\setlength{\abovecaptionskip}{-0pt}
	\setlength{\belowcaptionskip}{-15pt}
	\centering
	\includegraphics[width= 0.38\textwidth]{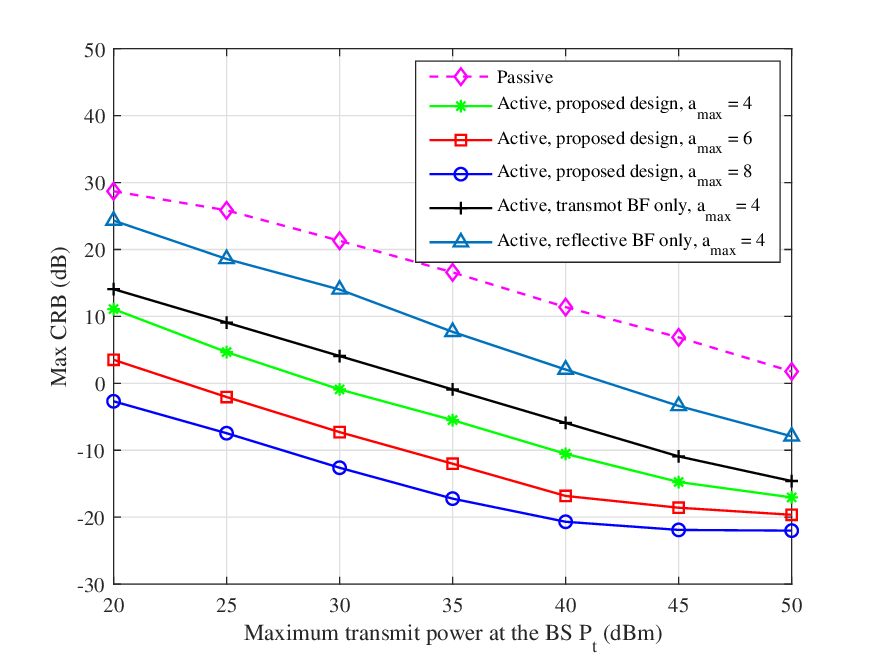}
	\DeclareGraphicsExtensions.
	\caption{\color{black}The achieved max-CRB versus the maximum transmit power $P_{\text{t}}$ at the BS with $P_{\text{s}} = 0.1$ W, $M=16$, and $N_v = N_h = 4$.}
	\label{CRBVsPt}
\end{figure}
Fig.~\ref{CRBVsPt} plots the achieved max-CRB versus the maximum transmit power $P_{\text{t}}$ at the BS. First, it is observed that our proposed design outperforms other benchmark schemes, and the max-CRB performance achieved by active IRSs outperforms that of passive IRSs by a significant margin. This clearly shows the benefit of deploying active IRSs for wireless sensing. Besides, the max-CRB achieved by the `Transmit BF only' benchmark is lower than that of the `Reflective BF only' one. This indicates that transmit beamforming plays a more prominent role in the considered sensing system. In particular, a delicate design of transmit beamforming to direct beams toward multiple active IRSs is a rule of thumb for establishing high-quality sensing links. Furthermore, it is also observed that the higher maximum amplification gain at the IRSs, the better sensing performance can be achieved. This is because a looser maximum amplification gain constraint of the elements at IRSs implies more degrees of freedom in reflective beamforming.

\begin{figure}[tbp]
	\setlength{\abovecaptionskip}{-0pt}
	\setlength{\belowcaptionskip}{-15pt}
	\centering
	\includegraphics[width= 0.38\textwidth]{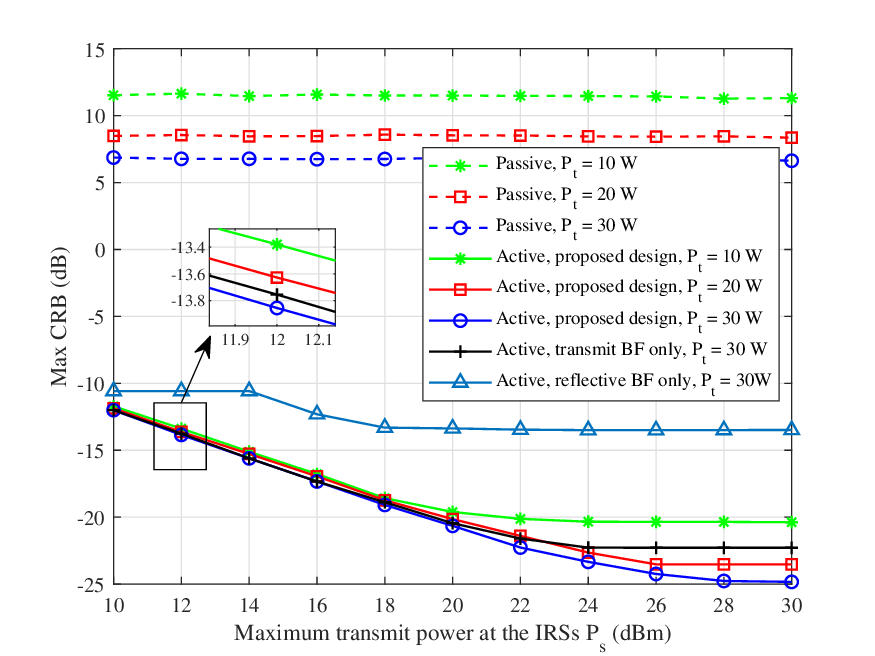}
	\DeclareGraphicsExtensions.
	\caption{\color{black}The achieved max-CRB versus the maximum transmit power $P_{\text{s}}$ at the IRSs with $M=16$, $N_v = N_h = 4$, and $a_{\text{max}}=8$.}
	\label{CRBVsPs}
\end{figure}
Fig.~\ref{CRBVsPs} shows the achieved max-CRB versus the maximum transmit power $P_{\text{s}}$ at the IRSs. In the low $P_{\text{s}}$ regime, it is observed that the CRBs under different $P_{\text{t}}$ are almost the same and decrease as $P_{\text{s}}$ increases. This phenomenon occurs because the received echo signal power is primarily constrained by the maximum transmit power budget at the IRSs. Subsequently, in the high $P_{\text{s}}$ regime, it is observed that the CRBs keep constants. This is attributed to the fact that the maximum transmit power budget at the IRS is large enough, and the received echo signal power is mainly constrained by the transmit power at the BS instead.

\begin{figure}[tbp]
	\setlength{\abovecaptionskip}{-0pt}
	\setlength{\belowcaptionskip}{-17pt}
	\centering
	\includegraphics[width= 0.38\textwidth]{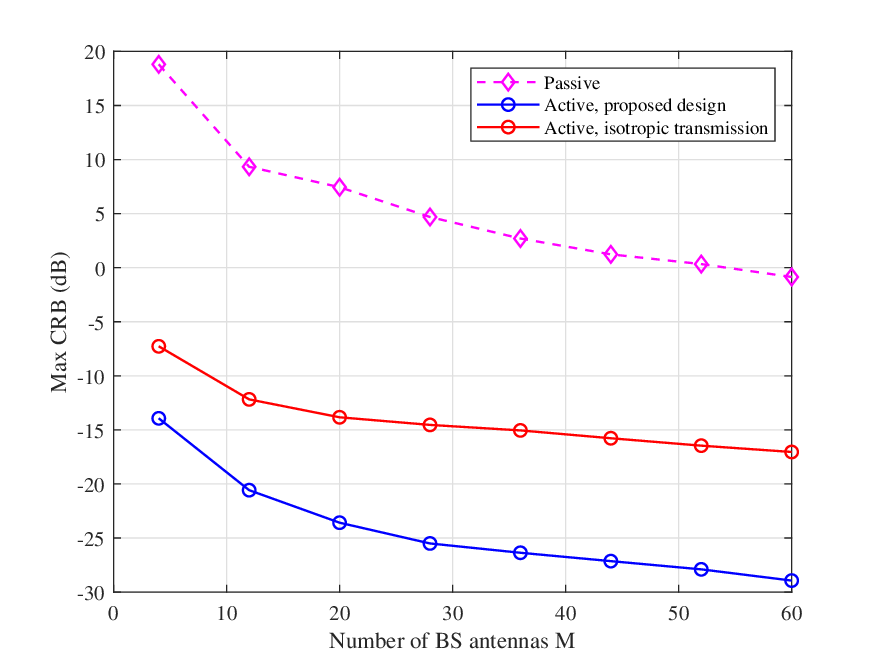}
	\DeclareGraphicsExtensions.
	\caption{\color{black}The achieved max-CRB versus the number of antennas $M$ at the BS with $P_{\text{t}} = 20$ W, $P_{\text{s}} = 0.1$ W $N_v = N_h = 4$, and $a_{\text{max}}=8$.}
	\label{CRBVsBSantenna}
\end{figure}
Fig.~\ref{CRBVsBSantenna} plots the achieved max-CRB versus the number of antennas $M$ at the BS. It is observed that the CRB decreases as  $M$ increases, but the performance saturates as $M$ further increases. In particular, when the number of antennas, $M$, starts to increase from a small number, the received echo signal power is enhanced rapidly due to the enlarged array gain. However, when $M$ further goes large, the array gain is no longer the limiting factor of sensing performance. Instead, the maximum transmit power budgets at the BS and IRSs limit the received echo signal power. This leads to a diminishing performance gain in terms of max-CRB when more antennas are deployed at the BS.


\section{Conclusion}
This paper investigated multi-active-IRS-assisted cooperative sensing, where multiple active IRSs are deployed to provide multi-view sensing. We proposed a multi-active-IRS time division sensing framework and then derived the closed-form CRBs for estimation of target parameters. Then, we proposed an efficient joint transmit and reflective beamforming design to minimize the maximum CRB among all IRSs, subject to the maximum transmit power budgets at the BS and IRSs, and the maximum amplification gain constraints at the IRSs. Numerical results demonstrated the effectiveness of our proposed design, and active IRSs outperform passive ones by a significant margin for target sensing. Furthermore,  it is shown that the maximum transmit power budget and the maximum amplification gain at the IRSs jointly limit the sensing performance, especially when the transmit power budget at the BS becomes large. Additionally, it indicates that the design of transmit beamforming is more critical than that of reflective beamforming.

\vspace{-2mm}
\begin{appendices}
	\section{Proof of Proposition \ref{prop_FIM_bs}}\label{prop_FIM1_proof}
	We define $\mathbf{D}_{N} = \mathrm{diag}\{\mathbf{d}_{N}\}$ with $\mathbf{d}_{N} = [0,1,\cdots,N-1]$, and then we derive the derivative of ${\mathbf{a}}_{l}$ with respect to $\theta_{l}$ as follows:
		\begin{align}
			&\dot{{\mathbf{a}}}_{\theta_{l}} = \frac{{\partial {{\mathbf{a}}_l}}}{{\partial {\theta _l}}} = \frac{{\partial {{{\mathbf{a}}}_v}({\theta _l}) \otimes {{{\mathbf{a}}}_h}({\theta _l},{\phi _l})}}{{\partial {\theta _l}}}\nonumber\\
			&=  - j\frac{{2\pi {{ d}_v}}}{\lambda }\sin ({\theta _l}){\mathbf{D}_{{{ N}_v}}}{{\mathbf{a}}_v}({\theta _l}) \otimes {{\mathbf{a}}_h}({\theta _l},{\phi _l}) \nonumber\\
			&+ j\frac{{2\pi {{ d}_h}}}{\lambda }\cos ({\theta _l})\cos ({\phi _l}){{\mathbf{a}}_v}({\theta _l}) \otimes {\mathbf{D}_{{{ N}_h}}}{{\mathbf{a}}_h}({\theta _l},{\phi _l})\nonumber\\
			&=- j\frac{{2\pi {{d}_v}}}{\lambda }\sin ({\theta _l})\left( {{\mathbf{d}_{{{ N}_v}}} \otimes {\mathbf{1}_{{{N}_h}}}} \right) \circ \left( {{{{\mathbf{a}}}_v}({\theta _l}) \otimes {{{\mathbf{a}}}_h}({\theta _l},{\phi _l})} \right)\nonumber\\  
			&+j\frac{{2\pi {{ d}_h}}}{\lambda }\cos ({\theta _l})\cos ({\phi _l})\left( {{\mathbf{1}_{{{ N}_v}}} \otimes {\mathbf{d}_{{{ N}_h}}}} \right) \circ \left( {{{{\mathbf{a}}}_v}({\theta _l}) \otimes {{{\mathbf{a}}}_h}({\theta _l},{\phi _l})} \right)\nonumber\\
			&= \left( j\frac{{2\pi {{ d}_h}}}{\lambda }\cos ({\theta _l})\cos ({\phi _l})\left( {{\mathbf{1}_{{{ N}_v}}} \otimes {\mathbf{d}_{{{ N}_h}}}} \right) \right.\nonumber\\
			&\left.- j\frac{{2\pi {{ d}_v}}}{\lambda }\sin ({\theta _l})\left( {{\mathbf{d}_{{{ N}_v}}} \otimes {\mathbf{1}_{{{ N}_h}}}} \right) \right) \circ {{\mathbf{a}}_l}\nonumber\\
			&= \bm{\zeta}_{\theta_l} \circ {{\mathbf{a}}_l} = \bm{Z}_{\theta_l} {{\mathbf{a}}_l},\label{De_a_theta}
		\end{align}
		where $\bm{Z}_{\theta_l} = \mathrm{diag}(\bm{\zeta}_{\theta_l})$ with $\bm{\zeta}_{\theta_l} = j\frac{{2\pi {{ d}_h}}}{\lambda }\cos ({\theta _l})\cos ({\phi _l})\left( {{\mathbf{1}_{{{ N}_v}}} \otimes {\mathbf{d}_{{{ N}_h}}}} \right) - j\frac{{2\pi {{ d}_v}}}{\lambda }\sin ({\theta _l})\left( {{\mathbf{d}_{{{ N}_v}}} \otimes {\mathbf{1}_{{{ N}_h}}}} \right) $ and $\mathbf{1}_{N} \in \mathbb{R}^{N\times1}$ with all elements are one. Similarly, the derivative of ${\mathbf{a}}_{l}$ with respect to $\phi_{l}$ is derived by
		\begin{align}
			&\dot{{\mathbf{a}}}_{\phi_{l}} = \frac{{\partial {{\mathbf{a}}_l}}}{{\partial {\phi_l}}} = \frac{{\partial {{{\mathbf{a}}}_v}({\theta _l}) \otimes {{{\mathbf{a}}}_h}({\theta _l},{\phi _l})}}{{\partial {\phi _l}}}\nonumber\\
			&= \frac{{\partial {{{\mathbf{a}}}_v}({\theta _l})}}{{\partial {\phi _l}}} \otimes {{\mathbf{a}}_h}({\theta _l},{\phi _l}) + {{\mathbf{a}}_v}({\theta _l}) \otimes \frac{{\partial {{{\mathbf{a}}}_h}({\theta _l},{\phi _l})}}{{ \partial {\phi _l}}}\nonumber\\
			&= \! - j\frac{{2\pi {{ d}_h}}}{\lambda }\sin ({\theta _l})\! \sin ({\phi _l})\! \left( {{\mathbf{1}_{{{ N}_v}}}\!\! \otimes {\mathbf{d}_{{{ N}_h}}}} \right)\! \circ \! \left( {{{{\mathbf{a}}}_v}({\theta _l})\!\! \otimes {{{\mathbf{a}}}_h}({\theta _l},{\phi _l})} \right)\nonumber\\
			&= \bm{\zeta}_{\phi_l} \circ {{\mathbf{a}}_l} = \bm{Z}_{\phi_l} {{\mathbf{a}}_l},\label{De_a_phi}
		\end{align}
		where $\bm{Z}_{\phi_l} = \mathrm{diag}(\bm{\zeta}_{\phi_l})$ with $\bm{\zeta}_{\phi_l} =- j\frac{{2\pi {{ d}_h}}}{\lambda }\sin ({\theta _l})\sin ({\phi _l})\left( {{\mathbf{1}_{{{ N}_v}}} \otimes {\mathbf{d}_{{{ N}_h}}}} \right)$. Next, we derive ${{F}}_{\theta_{l},\theta_l}$, ${{F}}_{\phi_{l},\phi_l}$, and ${\mathbf{F}}_{{\bm{\beta}_{l}},\bm{\beta}_{l}}$. According to the definition of FIM, ${{F}}_{\theta_{l},\theta_l}$ is given by 
			\begin{align}
				{{F}}_{\theta_{l},\theta_l} \!=\! \mathrm{tr}\left(\mathbf{R}_{{\mathbf{y}}_{l}}^{-1}\frac{\partial \mathbf{R}_{{\mathbf{y}}_{l}}}{\partial \theta_{l}}\mathbf{R}_{{\mathbf{y}}_{l}}^{-1}\frac{\partial \mathbf{R}_{{\mathbf{y}}_{l}}}{\partial \theta_{l}}\right) \!+\! 2\Re\left(\frac{\partial {\bm{\eta}}_{l}^{H}}{\partial \theta_{l} } \mathbf{R}_{{\mathbf{y}}_{l}}^{-1} \frac{\partial {\bm{\eta}}_{l}}{\partial \theta_{l} }\right). \label{FIM_bs_theta_theta}
			\end{align}
By substituting \eqref{R_y_bar_bs} into \eqref{FIM_bs_theta_theta}, we have
			\begin{align}
				&{\mathrm{tr}}\left( {{\mathbf{R}}_{{{ {\mathbf{y}} }_l}}^{ - 1}\frac{{\partial {{\mathbf{R}}_{{{ {\mathbf{y}} }_l}}}}}{{\partial {\theta _l}}}{\mathbf{R}}_{{{ {\mathbf{y}} }_l}}^{ - 1}\frac{{\partial {{\mathbf{R}}_{{{ {\mathbf{y}} }_l}}}}}{{\partial {\theta _l}}}} \right) = 0. \label{F_bs_bar_theta_theta_1_1}
			\end{align}
Then, substituting \eqref{vec_sig_bs} into \eqref{FIM_bs_theta_theta}, we obtain
{\small\begin{align}
	&\Re\left(\frac{\partial {\bm{\eta}}_{l}^{H}}{\partial \theta_{l} } \mathbf{R}_{{\mathbf{y}}_{l}}^{-1} \frac{\partial {\bm{\eta}}_{l}}{\partial \theta_{l} }\right) = \nonumber\\
	& \Re \left( {\sum\limits_{t = \frac{(l - 1)T_c}{L} + 1}^{l{\frac{T_c}{L}}}\!\!\!\!\!\!\! {{{\left( {\frac{{\partial \mathbf{G}_{l}^{T}\mathbf{\Psi}_{l}{\mathbf{E}}_{l}\mathbf{\Psi}_{l}\mathbf{G}_{l} {{\mathbf{s}}_l}[t]}}{{\partial {\theta _l}}}} \right)}^H}\!\!\!\!{\mathbf{R}_{{\mathbf{w}}_{l}}^{ - 1}}\frac{{\partial \mathbf{G}_{l}^{T}\mathbf{\Psi}_{l}{\mathbf{E}}_{l}\mathbf{\Psi}_{l}\mathbf{G}_{l}{{\mathbf{s}}_l}[t]}}{{\partial {\theta _l}}}} } \right)  \nonumber\\
	&= {\left| {{\beta _l}} \right|^2}\Re \left( \sum\limits_{t = \frac{(l - 1)T_c}{L} + 1}^{l{\frac{T_c}{L}}}  {\mathbf{s}}_l^H[t]{\mathbf{G}}_l^H{\mathbf{\Psi }}_l^H{{\left( {\dot{{\mathbf{a}}}_{\theta_{l}} {\mathbf{a}} _l^T + {{ {\mathbf{a}} }_l}{\dot{{\mathbf{a}}}_{\theta_{l}}^T}} \right)}^H} \right.\nonumber\\
	&\left.\mathbf{\Psi}_{l} \mathbf{G}_{l}^{*} {\mathbf{R}_{{\mathbf{w}}_{l}}^{ - 1}} \mathbf{G}_{l}^{T}\mathbf{\Psi}_{l} \left( {\dot{{\mathbf{a}}}_{\theta_{l}} {\mathbf{a}} _l^T + {{ {\mathbf{a}} }_l}{\dot{{\mathbf{a}}}_{\theta_{l}}^T}} \right){{\mathbf{\Psi }}_l}{{\mathbf{G}}_l}{{\mathbf{s}}_l}[t]  \right)\nonumber\\
	&={\left| {{\beta _l}} \right|^2}\frac{{{T_c}}}{L}\mathrm{tr}\left(  { \mathbf{C}_{\theta_{l},l}^H} {\mathbf{R}_{{\mathbf{w}}_{l}}^{ - 1}} \mathbf{C}_{\theta_{l},l} {\mathbf{R}_{s,l}} \right). \label{F_bs_bar_theta_theta_2}
\end{align}}where $\mathbf{C}_{\theta_{l},l} = \mathbf{G}_{l}^{T}\mathbf{\Psi}_{l} \left({\dot{{\mathbf{a}}}_{\theta_{l}} {\mathbf{a}} _l^T + {{ {\mathbf{a}} }_l}{\dot{{\mathbf{a}}}_{\theta_{l}}^T}}\right) {{\mathbf{\Psi }}_l}{{\mathbf{G}}_l}$. Based on \eqref{F_bs_bar_theta_theta_1_1} and \eqref{F_bs_bar_theta_theta_2}, ${\mathbf{F}}_{\theta_{l},\theta_l}$ is obtained.
Similarly, the closed-form expressions of ${{F}}_{\phi_{l},\phi_l}$, ${{F}}_{\theta_{l},\phi_l}$, ${{F}}_{\theta_{l},\bm{\beta}_{l}}$, ${{F}}_{\phi_{l},\bm{\beta}_{l}}$, and ${\mathbf{F}}_{{\bm{\beta}_{l}},\bm{\beta}_{l}}$ are given by
	\begin{align}
		&{{F}}_{\phi_{l},\phi_{l}} =  \frac{{{2 T_c}}}{L} {\left| {{\beta _l}} \right|^2}\mathrm{tr}\left(  {\mathbf{C}_{\phi_{l},l}^H}{\mathbf{R}_{{\mathbf{w}}_{l}}^{ - 1}}  {\mathbf{C}_{\phi_{l},l} }{\mathbf{R}_{s,l}} \right),\\
		&{ {{F}} _{{\theta_l},{\phi_l}}} =\frac{{{2 T_c}}}{L} {\left| {{\beta _l}} \right|^2}\mathrm{tr}\left(  {\mathbf{C}_{\theta_{l},l} ^H}  {\mathbf{R}_{{\mathbf{w}}_{l}}^{ - 1}}  {\mathbf{C}_{\phi_{l},l} } {\mathbf{R}_{s,l}} \right),\\
		&{\mathbf{F}}_{\theta_{l},{\bm{\beta}}_l} = \frac{{{2 T_c}}}{L}\Re \left( \beta_{l}\mathrm{tr}\left({\mathbf{C}_{\theta_{l},l} ^H} {\mathbf{R}_{{\mathbf{w}}_{l}}^{ - 1}}\mathbf{H}_{l} {\mathbf{R}_{s,l}} \right)[1, j]\right),\\
		&{\mathbf{F}}_{\phi_{l},\bm{\beta}_{l}} = \frac{{{2 T_c}}}{L}\Re \left( \beta_{l}\mathrm{tr}\left( {\mathbf{C}_{\phi_{l},l} ^H}  {\mathbf{R}_{{\mathbf{w}}_{l}}^{ - 1}}  \mathbf{H}_{l} {\mathbf{R}_{s,l}} \right)[1, j]\right),\\
		&{\mathbf{F}}_{{\bm{\beta}_{l}},\bm{\beta}_{l}} =  \frac{{{2 T_c}}}{L}\mathrm{tr}\left(  {\mathbf{H}_{l}^H}  {\mathbf{R}_{{\mathbf{w}}_{l}}^{ - 1}}  \mathbf{H}_{l} {\mathbf{R}_{s,l}} \right) \mathbf{I}_{2}, 
	\end{align}
respectively, where $\mathbf{C}_{\phi_{l},l} = \mathbf{G}_{l}^{T}\mathbf{\Psi}_{l} \left({\dot{{\mathbf{a}}}_{\phi_{l}} {\mathbf{a}} _l^T + {{ {\mathbf{a}} }_l}{\dot{{\mathbf{a}}}_{\phi_{l}}^T}}\right) {{\mathbf{\Psi }}_l}{{\mathbf{G}}_l}$ and $\mathbf{H}_{l} =\mathbf{G}_{l}^{T}\mathbf{\Psi}_{l} {{\mathbf{a}}}_{l} {\mathbf{a}} _l^T  {{\mathbf{\Psi }}_l}{{\mathbf{G}}_l}$ with $\dot{{\mathbf{a}}}_{\phi_{l}} = \frac{{\partial {{\mathbf{a}}_l}}}{{\partial {\phi _l}}}= \bm{Z}_{\phi_l} {{\mathbf{a}}_l}$. As a result, Proposition \ref{prop_FIM_bs} is proved.
	\section{Proof of Lemma \ref{FIM_appr_lem}}\label{FIM_appr_proof}
	Based on the partial derivative $\dot {\mathbf{a}}_{{\theta _l}} = \bm{Z}_{\theta_l} {{\mathbf{a}}_l}$ and $\dot {\mathbf{a}}_{{\phi _l}} = \bm{Z}_{\phi_l} {{\mathbf{a}}_l}$ in \eqref{De_a_theta} and \eqref{De_a_phi}, respectively, we have 
	\begin{align}
		\mathbf{C}_{\theta_{l},l} &= \mathbf{G}_{l}^{T} \mathbf{A}_{l}\left(\bm{Z}_{\theta_l} \bm{\psi}_{l}\bm{\psi}_{l}^{T}+\bm{\psi}_{l}\bm{\psi}_{l}^{T} \bm{Z}_{\theta_l} \right) \mathbf{A}_{l}{{\mathbf{G}}_l},\\
		\mathbf{C}_{\phi_{l},l} & = \mathbf{G}_{l}^{T} \mathbf{A}_{l}\left(\bm{Z}_{\phi_l} \bm{\psi}_{l}\bm{\psi}_{l}^{T}+\bm{\psi}_{l}\bm{\psi}_{l}^{T} \bm{Z}_{\phi_l} \right) \mathbf{A}_{l}{{\mathbf{G}}_l},\\
		\mathbf{H}_{l} & = \mathbf{G}_{l}^{T}\mathbf{A}_{l}\bm{\psi}_{l}{{\bm{\psi }}_l^{T}}\mathbf{A}_{l}{{\mathbf{G}}_l},
	\end{align}
	where $\mathbf{A}_{l} = \mathrm{diag}(\mathbf{a}_{l})$. By defining $\mathbf{\Theta}_{l} = \bm{\psi}_{l}\bm{\psi}_{l}^{H}$, we can transform the trace term in \eqref{F_theta_theta} as
	{\small\begin{align}
		&Q_{\theta_{l},\theta_l}(\mathbf{\Theta}_{l}) =\mathrm{tr}\left(  \mathbf{C}_{\theta_{l},l}^H {\mathbf{R}_{{\mathbf{w}}_{l}}^{ - 1}} \mathbf{C}_{\theta_{l},l} {\mathbf{R}_{s,l}}\right)\nonumber\\
	&={\mathrm{tr}}\left( \mathbf{G}_{l}^{H} \mathbf{A}_{l}^{H} \left(\bm{\psi}_{l}^{*} \bm{\psi}_{l}^{H} \bm{Z}_{\theta_l}^{H} + \bm{Z}_{\theta_l}^{H} \bm{\psi}_{l}^{*}\bm{\psi}_{l}^{H}  \right)\right.\nonumber\\
	& \left. \mathbf{A}_{l}^{H}{{\mathbf{G}}_l}^{*} {\mathbf{R}_{{\mathbf{w}}_{l}}^{ - 1}} \mathbf{G}_{l}^{T} \mathbf{A}_{l}\left(\bm{Z}_{\theta_l} \bm{\psi}_{l}\bm{\psi}_{l}^{T}+\bm{\psi}_{l}\bm{\psi}_{l}^{T} \bm{Z}_{\theta_l} \right) \mathbf{A}_{l}{{\mathbf{G}}_l} {\mathbf{R}_{s,l}} \right) \nonumber\\
	& ={\mathrm{tr}}\left( \mathbf{R}_{1} \mathbf{\Theta}_{l}^{T} \! \right) {\mathrm{tr}}\left(  \bm{Z}_{\theta_l}^{H}\mathbf{R}_{2}\bm{Z}_{\theta_l} \mathbf{\Theta}_{l} \! \right)\!\! +\! {\mathrm{tr}}\left( \bm{Z}_{\theta_l} \mathbf{R}_{1} \mathbf{\Theta}_{l}^{T}\! \right){\mathrm{tr}} \left(\bm{Z}_{\theta_l}^{H}\mathbf{R}_{2}\mathbf{\Theta}_{l}\right)\nonumber\\
	&+ {\mathrm{tr}}\left(\mathbf{R}_{1} \bm{Z}_{\theta_l}^{H}\mathbf{\Theta}_{l}^{T}\right) {\mathrm{tr}}\left(  \mathbf{R}_{2}\bm{Z}_{\theta_l}\mathbf{\Theta}_{l} \right)  + {\mathrm{tr}}\left( \bm{Z}_{\theta_l} \mathbf{R}_{1} \bm{Z}_{\theta_l}^{H} \mathbf{\Theta}_{l}^{T} \right){\mathrm{tr}} \left(  \mathbf{R}_{2} \mathbf{\Theta}_{l}\right),\label{F_theta_theta_trans}
	\end{align}}where $\mathbf{R}_{1} = \mathbf{A}_{l}{{\mathbf{G}}_l} {\mathbf{R}_{s,l}} \mathbf{G}_{l}^{H} \mathbf{A}_{l}^{H}$ and $\mathbf{R}_{2} = \mathbf{A}_{l}^{H}{{\mathbf{G}}_l}^{*} {\mathbf{R}_{{\mathbf{w}}_{l}}^{ - 1}} \mathbf{G}_{l}^{T} \mathbf{A}_{l}$. Then, we assume ${\mathbf{R}_{{\mathbf{w}}_{l}}^{ - 1}}$ is constant and derive the derivative of \eqref{F_theta_theta_trans} with respect to $\mathbf{\Theta}_{l}$ as 
	\begin{align}
		&\nabla_{\theta_{l},\theta_l}(\mathbf{\Theta}_{l})  = \frac{\partial}{\partial \mathbf{\Theta}_{l}}\mathrm{tr}\left(  \mathbf{C}_{\theta_{l},l}^H {\mathbf{R}_{{\mathbf{w}}_{l}}^{ - 1}} \mathbf{C}_{\theta_{l},l} {\mathbf{R}_{s,l}}\right)\nonumber\\
	& ={\mathrm{tr}}\left(  \bm{Z}_{\theta_l}^{H}\mathbf{R}_{2}\bm{Z}_{\theta_l} \mathbf{\Theta}_{l} \right)\mathbf{R}_{1} \!+\! {\mathrm{tr}}\left( \mathbf{R}_{1} \mathbf{\Theta}_{l}^{T}\right) \bm{Z}_{\theta_l}^{T}\mathbf{R}_{2}^{T}\bm{Z}_{\theta_l}^{*} \nonumber\\
	&+\! {\mathrm{tr}} \left(\bm{Z}_{\theta_l}^{H}\mathbf{R}_{2}\mathbf{\Theta}_{l}\right)\bm{Z}_{\theta_l} \mathbf{R}_{1} \! + \! {\mathrm{tr}}\left( \bm{Z}_{\theta_l} \mathbf{R}_{1} \mathbf{\Theta}_{l}^{T}\right)\mathbf{R}_{2}^{T} \bm{Z}_{\theta_l}^{*} \nonumber\\
	&+ {\mathrm{tr}}\left(  \mathbf{R}_{2}\bm{Z}_{\theta_l}\mathbf{\Theta}_{l} \right)\mathbf{R}_{1} \bm{Z}_{\theta_l}^{H} \!+\! {\mathrm{tr}}\left(\mathbf{R}_{1} \bm{Z}_{\theta_l}^{H}\mathbf{\Theta}_{l}^{T}\right)\bm{Z}_{\theta_l}^{T} \mathbf{R}_{2}^{T} \nonumber\\
	&+\! {\mathrm{tr}} \left(  \mathbf{R}_{2} \mathbf{\Theta}_{l}\right)\bm{Z}_{\theta_l} \mathbf{R}_{1} \bm{Z}_{\theta_l}^{H}  \!+\! {\mathrm{tr}}\left( \bm{Z}_{\theta_l} \mathbf{R}_{1} \bm{Z}_{\theta_l}^{H} \mathbf{\Theta}_{l}^{T} \right)\mathbf{R}_{2}^{T}.
	\end{align}
	As a result, the derivative of $F_{\theta_{l},\theta_l}$ is given by $\nabla{{F}}_{\theta_{l},\theta_{l}}(\mathbf{\Theta}_{l}^{(i)}) =  \frac{{{2 T_c}}}{L}{\left| {{\beta _l}} \right|^2} Q_{\theta_{l},\theta_l}$. Similar to \eqref{F_theta_theta_trans}, the trace terms in \eqref{F_phi_phi}-\eqref{F_phi_beta} are transformed into 
	\begin{align}
	&Q_{\phi_{l},\phi_l}(\mathbf{\Theta}_{l}) = \mathrm{tr}\left(  \mathbf{C}_{\phi_{l},l}^H {\mathbf{R}_{{\mathbf{w}}_{l}}^{ - 1}} \mathbf{C}_{\phi_{l},l} {\mathbf{R}_{s,l}}\right) \nonumber\\
	&={\mathrm{tr}}\left( \mathbf{R}_{1} \mathbf{\Theta}_{l}^{T}\right) {\mathrm{tr}}\left(  \bm{Z}_{\phi_l}^{H}\mathbf{R}_{2}\bm{Z}_{\phi_l} \mathbf{\Theta}_{l} \right) \nonumber\\
	&+ {\mathrm{tr}}\left( \bm{Z}_{\phi_l} \mathbf{R}_{1} \mathbf{\Theta}_{l}^{T}\right){\mathrm{tr}} \left(\bm{Z}_{\phi_l}^{H}\mathbf{R}_{2}\mathbf{\Theta}_{l}\right)\nonumber\\
	&+ {\mathrm{tr}}\left(\mathbf{R}_{1} \bm{Z}_{\phi_l}^{H}\mathbf{\Theta}_{l}^{T}\right) {\mathrm{tr}}\left(  \mathbf{R}_{2}\bm{Z}_{\phi_l}\mathbf{\Theta}_{l} \right) \nonumber\\
	& + {\mathrm{tr}}\left( \bm{Z}_{\phi_l} \mathbf{R}_{1} \bm{Z}_{\phi_l}^{H} \mathbf{\Theta}_{l}^{T} \right){\mathrm{tr}} \left(  \mathbf{R}_{2} \mathbf{\Theta}_{l}\right),\label{F_phi_phi_trans}\\
	&Q_{\theta_{l},\phi_l}(\mathbf{\Theta}_{l}) = \mathrm{tr}\left(  \mathbf{C}_{\theta_{l},l}^H {\mathbf{R}_{{\mathbf{w}}_{l}}^{ - 1}} \mathbf{C}_{\phi_{l},l} {\mathbf{R}_{s,l}}\right) \nonumber\\
	&={\mathrm{tr}}\left( \mathbf{R}_{1} \mathbf{\Theta}_{l}^{T}\right) {\mathrm{tr}}\left(  \bm{Z}_{\theta_l}^{H}\mathbf{R}_{2}\bm{Z}_{\phi_l} \mathbf{\Theta}_{l} \right) \nonumber\\
	&+ {\mathrm{tr}}\left( \bm{Z}_{\phi_l} \mathbf{R}_{1} \mathbf{\Theta}_{l}^{T}\right){\mathrm{tr}} \left(\bm{Z}_{\theta_l}^{H}\mathbf{R}_{2}\mathbf{\Theta}_{l}\right)\nonumber\\
	&+ {\mathrm{tr}}\left(\mathbf{R}_{1} \bm{Z}_{\theta_l}^{H}\mathbf{\Theta}_{l}^{T}\right) {\mathrm{tr}}\left(  \mathbf{R}_{2}\bm{Z}_{\phi_l}\mathbf{\Theta}_{l} \right) \nonumber\\
	&+ {\mathrm{tr}}\left( \bm{Z}_{\phi_l} \mathbf{R}_{1} \bm{Z}_{\theta_l}^{H} \mathbf{\Theta}_{l}^{T} \right){\mathrm{tr}} \left(  \mathbf{R}_{2} \mathbf{\Theta}_{l}\right),\label{F_theta_phi_trans} \\
	&Q_{\theta_{l},\bm{\beta}_{l}}(\mathbf{\Theta}_{l}) = \mathrm{tr}\left( \mathbf{C}_{\theta_{l},l}^H  {\mathbf{R}_{{\mathbf{w}}_{l}}^{ - 1}}\mathbf{H}_{l} {\mathbf{R}_{s,l}} \right) =\nonumber\\
	& {\mathrm{tr}}\left(\mathbf{R}_{1} \mathbf{\Theta}_{l}^{T}\right)\! {\mathrm{tr}}\left(\bm{Z}_{\theta_l}^{H} \mathbf{R}_{2}\mathbf{\Theta}_{l} \right) \!+\! {\mathrm{tr}}\left(\mathbf{R}_{1} \bm{Z}_{\theta_l}^{H} \mathbf{\Theta}_{l}^{T}\right)\! {\mathrm{tr}}\left( \mathbf{R}_{2}\mathbf{\Theta}_{l} \right)\!, \label{F_theta_beta_trans} \\
	&Q_{\phi_{l},\bm{\beta}_{l}}(\mathbf{\Theta}_{l})  = \mathrm{tr}\left(\mathbf{C}_{\phi_{l},l}^H   {\mathbf{R}_{{\mathbf{w}}_{l}}^{ - 1}} \mathbf{H}_{l} {\mathbf{R}_{s,l}} \right)= \nonumber\\
	& {\mathrm{tr}}\left(\mathbf{R}_{1} \mathbf{\Theta}_{l}^{T}\right)\! {\mathrm{tr}}\left(\bm{Z}_{\phi_l}^{H} \mathbf{R}_{2}\mathbf{\Theta}_{l} \right) \!+\! {\mathrm{tr}}\left(\mathbf{R}_{1} \bm{Z}_{\phi_l}^{H} \mathbf{\Theta}_{l}^{T}\right)\! {\mathrm{tr}}\left( \mathbf{R}_{2}\mathbf{\Theta}_{l} \right)\!, \label{F_phi_beta_trans}\\
	&Q_{{\bm{\beta}_{l}},\bm{\beta}_{l}}(\mathbf{\Theta}_{l}) \!=\! \mathrm{tr}\left(  \mathbf{H}_{l}^H   {\mathbf{R}_{{\mathbf{w}}_{l}}^{ - 1}} \mathbf{H}_{l} {\mathbf{R}_{s,l}} \right) \!=\! {\mathrm{tr}}\left(\mathbf{R}_{1} \mathbf{\Theta}_{l}^{T}\right)\! {\mathrm{tr}}\left(\mathbf{R}_{2}\mathbf{\Theta}_{l} \right), \label{F_beta_beta_trans}
	\end{align}
respectively, and the corresponding derivatives of them with respect to $\mathbf{\Theta}_{l}$ are given by 
\begin{align}
	&\nabla_{\phi_{l},\phi_l}(\mathbf{\Theta}_{l}) = \frac{\partial}{\partial \mathbf{\Theta}_{l}} \mathrm{tr}\left(  \mathbf{C}_{\phi_{l},l}^H {\mathbf{R}_{{\mathbf{w}}_{l}}^{ - 1}} \mathbf{C}_{\phi_{l},l} {\mathbf{R}_{s,l}}\right) \nonumber\\
	&={\mathrm{tr}}\left(  \bm{Z}_{\phi_l}^{H}\mathbf{R}_{2}\bm{Z}_{\phi_l} \mathbf{\Theta}_{l} \right)\mathbf{R}_{1} + {\mathrm{tr}}\left( \mathbf{R}_{1} \mathbf{\Theta}_{l}^{T}\right) \bm{Z}_{\phi_l}^{T}\mathbf{R}_{2}^{T}\bm{Z}_{\phi_l}^{*} \nonumber\\
	&+ {\mathrm{tr}} \left(\bm{Z}_{\phi_l}^{H}\mathbf{R}_{2}\mathbf{\Theta}_{l}\right)\bm{Z}_{\phi_l} \mathbf{R}_{1}  + {\mathrm{tr}}\left( \bm{Z}_{\phi_l} \mathbf{R}_{1} \mathbf{\Theta}_{l}^{T}\right)\mathbf{R}_{2}^{T} \bm{Z}_{\phi_l}^{*} \nonumber\\
	&+ {\mathrm{tr}}\left(  \mathbf{R}_{2}\bm{Z}_{\phi_l}\mathbf{\Theta}_{l} \right)\mathbf{R}_{1} \bm{Z}_{\phi_l}^{H} + {\mathrm{tr}}\left(\mathbf{R}_{1} \bm{Z}_{\phi_l}^{H}\mathbf{\Theta}_{l}^{T}\right)\bm{Z}_{\phi_l}^{T} \mathbf{R}_{2}^{T} \nonumber\\
	&+ {\mathrm{tr}} \left(  \mathbf{R}_{2} \mathbf{\Theta}_{l}\right)\bm{Z}_{\phi_l} \mathbf{R}_{1} \bm{Z}_{\phi_l}^{H}  + {\mathrm{tr}}\left( \bm{Z}_{\phi_l} \mathbf{R}_{1} \bm{Z}_{\phi_l}^{H} \mathbf{\Theta}_{l}^{T} \right)\mathbf{R}_{2}^{T},\\	
	&\nabla_{\theta_{l},\phi_l}(\mathbf{\Theta}_{l}) = \frac{\partial}{\partial \mathbf{\Theta}_{l}} \mathrm{tr}\left(  \mathbf{C}_{\theta_{l},l}^H {\mathbf{R}_{{\mathbf{w}}_{l}}^{ - 1}} \mathbf{C}_{\phi_{l},l} {\mathbf{R}_{s,l}}\right) \nonumber\\
	&={\mathrm{tr}}\left(  \bm{Z}_{\theta_l}^{H}\mathbf{R}_{2}\bm{Z}_{\phi_l} \mathbf{\Theta}_{l} \right)\mathbf{R}_{1} + {\mathrm{tr}}\left( \mathbf{R}_{1} \mathbf{\Theta}_{l}^{T}\right) \bm{Z}_{\phi_l}^{T}\mathbf{R}_{2}^{T}\bm{Z}_{\theta_l}^{*} \nonumber\\
	&+ {\mathrm{tr}} \left(\bm{Z}_{\theta_l}^{H}\mathbf{R}_{2}\mathbf{\Theta}_{l}\right)\bm{Z}_{\phi_l} \mathbf{R}_{1}  + {\mathrm{tr}}\left( \bm{Z}_{\phi_l} \mathbf{R}_{1} \mathbf{\Theta}_{l}^{T}\right)\mathbf{R}_{2}^{T} \bm{Z}_{\theta_l}^{*} \nonumber\\
	&+ {\mathrm{tr}}\left(  \mathbf{R}_{2}\bm{Z}_{\phi_l}\mathbf{\Theta}_{l} \right)\mathbf{R}_{1} \bm{Z}_{\theta_l}^{H} + {\mathrm{tr}}\left(\mathbf{R}_{1} \bm{Z}_{\theta_l}^{H}\mathbf{\Theta}_{l}^{T}\right)\bm{Z}_{\phi_l}^{T} \mathbf{R}_{2}^{T} \nonumber\\
	&+ {\mathrm{tr}} \left(  \mathbf{R}_{2} \mathbf{\Theta}_{l}\right)\bm{Z}_{\phi_l} \mathbf{R}_{1} \bm{Z}_{\theta_l}^{H}  + {\mathrm{tr}}\left( \bm{Z}_{\phi_l} \mathbf{R}_{1} \bm{Z}_{\theta_l}^{H} \mathbf{\Theta}_{l}^{T} \right)\mathbf{R}_{2}^{T},
\end{align}	
\begin{align}
	&\nabla_{\theta_{l},\bm{\beta}_{l}}(\mathbf{\Theta}_{l})  = \frac{\partial}{\partial \mathbf{\Theta}_{l}} \mathrm{tr}\left( \mathbf{C}_{\theta_{l},l}^H  {\mathbf{R}_{{\mathbf{w}}_{l}}^{ - 1}}\mathbf{H}_{l} {\mathbf{R}_{s,l}} \right) \nonumber\\
	&= {\mathrm{tr}}\left(\bm{Z}_{\theta_l}^{H} \mathbf{R}_{2}\mathbf{\Theta}_{l} \right)\mathbf{R}_{1}  + {\mathrm{tr}}\left(\mathbf{R}_{1} \mathbf{\Theta}_{l}^{T}\right) \mathbf{R}_{2}^{T} \bm{Z}_{\theta_l}^{*}\nonumber\\
	&+ {\mathrm{tr}}\left( \mathbf{R}_{2}\mathbf{\Theta}_{l} \right)\mathbf{R}_{1} \bm{Z}_{\theta_l}^{H}  + {\mathrm{tr}}\left(\mathbf{R}_{1} \bm{Z}_{\theta_l}^{H} \mathbf{\Theta}_{l}^{T}\right) \mathbf{R}_{2}^{T},\\
	&\nabla_{\phi_{l},\bm{\beta}_{l}}(\mathbf{\Theta}_{l})  = \frac{\partial}{\partial \mathbf{\Theta}_{l}} \mathrm{tr}\left( \mathbf{C}_{\phi_{l},l}^H  {\mathbf{R}_{{\mathbf{w}}_{l}}^{ - 1}}\mathbf{H}_{l} {\mathbf{R}_{s,l}} \right) \nonumber\\
	&= {\mathrm{tr}}\left(\bm{Z}_{\phi_l}^{H} \mathbf{R}_{2}\mathbf{\Theta}_{l} \right)\mathbf{R}_{1}  + {\mathrm{tr}}\left(\mathbf{R}_{1} \mathbf{\Theta}_{l}^{T}\right) \mathbf{R}_{2}^{T} \bm{Z}_{\phi_l}^{*}\nonumber\\
	&+ {\mathrm{tr}}\left( \mathbf{R}_{2}\mathbf{\Theta}_{l} \right)\mathbf{R}_{1} \bm{Z}_{\phi_l}^{H}  + {\mathrm{tr}}\left(\mathbf{R}_{1} \bm{Z}_{\phi_l}^{H} \mathbf{\Theta}_{l}^{T}\right) \mathbf{R}_{2}^{T},\\
	&\nabla_{{\bm{\beta}_{l}},\bm{\beta}_{l}}(\mathbf{\Theta}_{l}) = \frac{\partial}{\partial \mathbf{\Theta}_{l}} \mathrm{tr}\left(  \mathbf{H}_{l}^H   {\mathbf{R}_{{\mathbf{w}}_{l}}^{ - 1}} \mathbf{H}_{l} {\mathbf{R}_{s,l}} \right) \nonumber\\
	&= {\mathrm{tr}}\left(\mathbf{R}_{2}\mathbf{\Theta}_{l} \right)\mathbf{R}_{1} + {\mathrm{tr}}\left(\mathbf{R}_{1} \mathbf{\Theta}_{l}^{T}\right) \mathbf{R}_{2}^{T}.
\end{align} 
As a result, Lemma \ref{FIM_appr_lem} is proved.
	\end{appendices}
	\bibliographystyle{IEEEtran}
	\bibliography{IEEEabrv,myref}

\begin{thebibliography}{10}
\providecommand{\url}[1]{#1}
\csname url@samestyle\endcsname
\providecommand{\newblock}{\relax}
\providecommand{\bibinfo}[2]{#2}
\providecommand{\BIBentrySTDinterwordspacing}{\spaceskip=0pt\relax}
\providecommand{\BIBentryALTinterwordstretchfactor}{4}
\providecommand{\BIBentryALTinterwordspacing}{\spaceskip=\fontdimen2\font plus
\BIBentryALTinterwordstretchfactor\fontdimen3\font minus
  \fontdimen4\font\relax}
\providecommand{\BIBforeignlanguage}[2]{{%
\expandafter\ifx\csname l@#1\endcsname\relax
\typeout{** WARNING: IEEEtran.bst: No hyphenation pattern has been}%
\typeout{** loaded for the language `#1'. Using the pattern for}%
\typeout{** the default language instead.}%
\else
\language=\csname l@#1\endcsname
\fi
#2}}
\providecommand{\BIBdecl}{\relax}
\BIBdecl

\bibitem{cui2021integrating}
Y.~Cui, F.~Liu, X.~Jing, and J.~Mu, ``{Integrating sensing and communications
  for ubiquitous IoT: Applications, trends, and challenges},'' \emph{IEEE
  Netw.}, vol.~35, no.~5, pp. 158--167, Sep. 2021.

\bibitem{liu2022integrated}
F.~Liu, Y.~Cui, C.~Masouros, J.~Xu, T.~X. Han, Y.~C. Eldar, and S.~Buzzi,
  ``{Integrated sensing and communications: Towards dual-functional wireless
  networks for 6G and beyond},'' \emph{IEEE J. Sel. Areas Commun.}, vol.~40,
  no.~6, pp. 1728--1767, Jun. 2022.

\bibitem{huang2019reconfigurable}
C.~Huang, A.~Zappone, G.~C. Alexandropoulos, M.~Debbah, and C.~Yuen,
  ``{Reconfigurable intelligent surfaces for energy efficiency in wireless
  communication},'' \emph{IEEE Trans. Wireless Commun.}, vol.~18, no.~8, pp.
  4157--4170, Aug. 2019.

\bibitem{wu2021intelligent}
Q.~Wu, S.~Zhang, B.~Zheng, C.~You, and R.~Zhang, ``{Intelligent reflecting
  surface-aided wireless communications: A tutorial},'' \emph{IEEE Trans.
  Commun.}, vol.~69, no.~5, pp. 3313--3351, May 2021.

\bibitem{song2023intelligent}
X.~Song, J.~Xu, F.~Liu, T.~X. Han, and Y.~C. Eldar, ``{Intelligent reflecting
  surface enabled sensing: Cram{\'e}r-Rao bound optimization},'' \emph{IEEE
  Trans. Signal Process.}, vol.~71, pp. 2011--2026, May 2023.

\bibitem{fang2023multiirsenabled}
Y.~Fang, S.~Zhang, X.~Li, X.~Yu, J.~Xu, and S.~Cui, ``{Multi-IRS-enabled
  integrated sensing and communications},'' \emph{IEEE Trans. Commun., Early
  Access}, 2024.

\bibitem{song2024cramer}
X.~Song, X.~Qin, J.~Xu, and R.~Zhang, ``{Cram{\'e}r-rao bound minimization for
  IRS-enabled multiuser integrated sensing and communications},'' \emph{IEEE
  Trans. Wireless Commun., Early Access}, 2024.

\bibitem{buzzi2022foundations}
S.~Buzzi, E.~Grossi, M.~Lops, and L.~Venturino, ``{Foundations of MIMO radar
  detection aided by reconfigurable intelligent surfaces},'' \emph{IEEE Trans.
  Signal Process.}, vol.~70, pp. 1749--1763, Mar. 2022.

\bibitem{zhang2022active}
Z.~Zhang, L.~Dai, X.~Chen, C.~Liu, F.~Yang, R.~Schober, and H.~V. Poor,
  ``{Active RIS vs. passive RIS: Which will prevail in 6G?}'' \emph{IEEE Trans.
  Commun.}, vol.~71, no.~3, pp. 1707--1725, Mar. 2023.

\bibitem{kang2024active}
Z.~Kang, C.~You, and R.~Zhang, ``{Active-IRS-aided wireless communication:
  Fundamentals, designs and open issues},'' \emph{IEEE Wireless Commun., Early
  Access}, 2024.

\bibitem{salem2022active}
A.~A. Salem, M.~H. Ismail, and A.~S. Ibrahim, ``{Active reconfigurable
  intelligent surface-assisted MISO integrated sensing and communication
  systems for secure operation},'' \emph{IEEE Trans. Veh. Technol.}, vol.~72,
  no.~4, pp. 4919--4931, Apr. 2023.

\bibitem{zhang2022CRAN}
Y.~Zhang, J.~Chen, C.~Zhong, H.~Peng, and W.~Lu, ``{Active IRS-assisted
  integrated sensing and communication in C-RAN},'' \emph{IEEE Wireless Commun.
  Lett.}, vol.~12, no.~3, pp. 411--415, Mar. 2023.

\bibitem{zhu2023joint}
Q.~Zhu, M.~Li, R.~Liu, and Q.~Liu, ``{Joint transceiver beamforming and
  reflecting design for active RIS-aided ISAC systems},'' \emph{IEEE Trans.
  Veh. Technol.}, vol.~72, no.~7, pp. 9636--9640, Jul. 2023.

\bibitem{zheng2022survey}
B.~Zheng, C.~You, W.~Mei, and R.~Zhang, ``A survey on channel estimation and
  practical passive beamforming design for intelligent reflecting surface aided
  wireless communications,'' \emph{IEEE Commun. Surv. Tuts.}, vol.~24, no.~2,
  pp. 1035--1071, Feb. 2022.

\bibitem{gavish1992performance}
M.~Gavish and A.~J. Weiss, ``{Performance analysis of bearing-only target
  location algorithms},'' \emph{IEEE Trans. Aerosp. Electron. Sys.}, vol.~28,
  no.~3, pp. 817--828, Jul. 1992.

\bibitem{kay1993fundamentals}
S.~M. Kay, \emph{{Fundamentals of Statistical Signal Processing: Estimation
  Theory}}.\hskip 1em plus 0.5em minus 0.4em\relax Cliffs, NJ, USA:
  Prentice-Hall., 1993.

\bibitem{grant2014cvx}
M.~Grant and S.~Boyd, ``{CVX: Matlab software for disciplined convex
  programming, version 2.1},'' 2014.

\bibitem{luo2010semidefinite}
Z.-Q. Luo, W.-K. Ma, A.~M.-C. So, Y.~Ye, and S.~Zhang, ``Semidefinite
  relaxation of quadratic optimization problems,'' \emph{IEEE Signal Process.
  Mag.}, vol.~27, no.~3, pp. 20--34, May 2010.

\bibitem{shao2022target}
X.~Shao, C.~You, W.~Ma, X.~Chen, and R.~Zhang, ``{Target sensing with
  intelligent reflecting surface: Architecture and performance},'' \emph{IEEE
  J. Sel. Areas Commun.}, vol.~40, no.~7, pp. 2070--2084, Jul. 2022.

\end{thebibliography}

\end{document}